\documentclass[11pt,a4paper]{article}
\usepackage{amsmath}
\usepackage{amsfonts,amssymb}
\usepackage[dvips]{graphics}
\begin{document}

\newtheorem{theorem}{Theorem}[section]
\newtheorem{lemma}{Lemma}[section]
\newtheorem{conjecture}{Conjecture}[section]
\newcommand{\bproof}{\smallskip{\em Proof. }}
\newcommand{\eproof}{\smallskip}
\newcommand{\qed}{{\hbox{$\  \square$}}}

\newcommand{\nl}{\nonumber\\}
\newcommand{\nnl}{\nl[6mm]}
\newcommand{\nle}{\nl[-2.3mm]\\[-2.3mm]}
\newcommand{\nlb}[1]{\nl[-2.3mm]\label{#1}\\[-2.3mm]}

\renewcommand{\theequation}{\thesection.\arabic{equation}}
\let\ssection=\section
\renewcommand{\section}{\setcounter{equation}{0}\ssection}

\newcommand{\be}{\bes}
\newcommand{\ee}{\ees}
\newcommand{\bes}{\begin{eqnarray}}
\newcommand{\ees}{\end{eqnarray}}
\newcommand{\eens}{\nonumber\end{eqnarray}}

\newcommand{\softbreak}{\break}

\renewcommand{\/}{\over}
\renewcommand{\d}{\partial}
\renewcommand{\div}{\hbox{div}}
\newcommand{\im}{\hbox{im\,}}

\newcommand{\wwedge}{\!\wedge\!}
\newcommand{\ootimes}{\!\otimes\!}
\newcommand{\ooplus}{\!\oplus\!}
\newcommand{\stimes}{\boxtimes}

\newcommand{\eps}{\epsilon}
\newcommand{\vth}{\vartheta}
\renewcommand{\th}{\theta}
\newcommand{\dlt}{\delta}
\newcommand{\ups}{\upsilon}
\newcommand{\al}{\alpha}
\newcommand{\bt}{\beta}
\newcommand{\gm}{\gamma}
\newcommand{\om}{\omega}
\newcommand{\Om}{\Omega}
\newcommand{\si}{\sigma}
\newcommand{\la}{\lambda}
\newcommand{\rep}{\varrho}

\newcommand{\du}{du}
\newcommand{\dz}{dz}
\newcommand{\dt}{dt}
\newcommand{\dth}{d\th}
\newcommand{\dvth}{d\vth}

\newcommand{\yj}{\eta^j}
\newcommand{\yk}{\eta^k}

\newcommand{\tQ}{\widetilde Q}
\newcommand{\tR}{\widetilde R}
\newcommand{\toj}{\tilde\oj}
\newcommand{\boj}{\bar\oj}

\newcommand{\LL}{{\cal L}}
\newcommand{\J}{{\cal J}}
\renewcommand{\S}{{\cal S}}
\newcommand{\G}{{\cal G}}

\newcommand{\symm}{\mathbf{S}}

\newcommand{\oj}{{\mathfrak g}}
\newcommand{\tr}{\hbox{tr}\,}
\newcommand{\half}{\hbox{$1\/2$}}
\newcommand{\mhalf}{\hbox{$1\/2$}(-)^X}
\newcommand{\third}{\hbox{$1\/3$}}
\newcommand{\mthird}{\hbox{$1\/3$}(-)^X}
\newcommand{\fourth}{\hbox{$1\/4$}}
\newcommand{\mfourth}{\hbox{$1\/4$}(-)^X}
\newcommand{\fifth}{\hbox{$1\/5$}}
\newcommand{\sixth}{\hbox{$1\/6$}}
\newcommand{\msixth}{\hbox{$1\/6$}(-)^X}
\newcommand{\eighth}{\hbox{$1\/8$}}

\newcommand{\ssg}{sl(3)\ooplus sl(2)\ooplus gl(1)}
\newcommand{\vect}{{\mathfrak{vect}}}
\newcommand{\svect}{{\mathfrak{svect}}}
\newcommand{\map}{{\mathfrak{map}}}
\newcommand{\mb}{{\mathfrak{mb}}}
\newcommand{\ksle}{{\mathfrak{ksle}}}
\newcommand{\bigksle}{\overline{{\mathfrak{ksle}}(5|10)}}
\newcommand{\vle}{{\mathfrak{vle}}}
\newcommand{\kas}{{\mathfrak{kas}}}
\newcommand{\vas}{{\mathfrak{vas}}}
\newcommand{\as}{{\mathfrak{as}}}
\newcommand{\kk}{{\mathfrak{k}}}
\newcommand{\fle}{{\mathfrak {le}}}
\newcommand{\sle}{{\mathfrak {sle}}}
\newcommand{\ko}{{\mathfrak {ko}}}
\newcommand{\sko}{{\mathfrak {sko}}}

\newcommand{\brep}[1]{{\bf{{#1}}}}

\newcommand{\barr}{\begin{array}}
\newcommand{\earr}{\end{array}}

\newcommand{\nm}{{n|m}}

\newcommand{\into}{\hookrightarrow}
\newcommand{\onto}{\twoheadrightarrow}
\newcommand{\hbreak}{\hfill\break}
\newcommand{\larroww}[1]{{\ \stackrel{#1}{\longleftarrow}\ }}
\newcommand{\rarroww}[1]{{\ \stackrel{#1}{\longrightarrow}\ }}

\newcommand{\vectmapg}{{\vect(N)\ltimes\map(N,\oj)}}
\newcommand{\vectmapv}{{\vect(N)\ltimes\map(N,\vect(\nm))}}

\newcommand{\ud}{u^i\d_i}
\newcommand{\thd}{\th_{ia}d^{ia}}
\newcommand{\veth}{\vth^a\eth_a}
\newcommand{\intth}{\hbox{$\int$}_\th}
\newcommand{\intu}{\hbox{$\int$}_u}
\newcommand{\intvth}{\hbox{$\int$}_\vth}

\newcommand{\ip}{{i_1..i_p}}
\newcommand{\iq}{{i_1..i_q}}
\newcommand{\is}{{i_1..i_s}}
\newcommand{\jq}{{j_1..j_q}}
\newcommand{\js}{{j_1..j_s}}
\newcommand{\klq}{{k_1l_1..k_ql_q}}
\newcommand{\jkr}{{j_1k_1..j_rk_r}}
\newcommand{\mnr}{{m_1n_1..m_rn_r}}
\newcommand{\ar}{{a_1..a_r}}

\newcommand{\NN}{{\mathbb N}}
\newcommand{\CC}{{\mathbb C}}
\newcommand{\RR}{{\mathbb R}}
\newcommand{\ZZ}{{\mathbb Z}}

\newcommand{\bal}{\bar\alpha}
\newcommand{\bbt}{\bar\beta}
\newcommand{\bgm}{\bar\gamma}
\newcommand{\bpsi}{\bar\psi}
\newcommand{\bchi}{\bar\chi}
\newcommand{\bla}{\bar\la}
\newcommand{\bv}{\bar v}

\newcommand{\DD}{{\cal D}}
\newcommand{\Dslash}{\hbox{$\DD\kern-2.4mm/$}}

\newcommand{\Lxi}{\LL_\xi}
\newcommand{\LX}{\LL_X}
\newcommand{\LY}{\LL_Y}
\newcommand{\LXY}{\LL_{[X,Y]}}
\newcommand{\LZ}{\LL_Z}
\newcommand{\xmu}{\xi^\mu}
\newcommand{\xnu}{\xi^\nu}
\newcommand{\ynu}{\eta^\nu}
\newcommand{\Dmu}{\DD_\mu}
\newcommand{\dmu}{\d_\mu}
\newcommand{\dnu}{\d_\nu}

\newcommand{\duth}{d^3u d^6\th\,}
\newcommand{\duthh}{d^5u d^{10}\th\,}
\newcommand{\duthv}{d^3u d^6\th d^2\vth\,}
\newcommand{\dNx}{d^N\!x\,}
\newcommand{\dnmy}{d^\nm\!y\,}

\title{{Exceptional Lie Superalgebras, Invariant Morphisms, 
and a Second-Gauged Standard Model}}

\author{T. A. Larsson \\
Vanadisv\"agen 29, S-113 23 Stockholm, Sweden,\\
 email: Thomas.Larsson@hdd.se }

\maketitle
\begin{abstract}
Degenerate modules of the exceptional infinite-dimensional simple Lie 
superalgebras $\vle(3|6)$, $\ksle(5|10)$ and $\mb(3|8)$
have recently been constructed
by Kac and Rudakov, and by Grozman, Leites and Shchepochkina.
I rederive their results using a formalism which is contragredient to
theirs; instead of finding singular vectors in induced modules, I build
reducible tensor modules (``forms'') from elementary differentials.
There is a discrepancy between my result for $\ksle(5|10)$ and Kac' and
Rudakov's one.

Since the grade zero subalgebra of $\vle(3|6)$ and $\mb(3|8)$ is
$\ssg$, gauge theories based on these algebras can be viewed as a
``second-gauged'' version of the standard model, where the rigid
$\ssg$ symmetry is made local not only in spacetime (``first gauging''),
but in the internal directions as well. An attempt to construct such
a second-gauged theory is presented. Some predictions regarding the
fermion spectrum, absense of new gauge bosons, and CP violation follow
immediately.
\end{abstract}

\newpage

\section{Introduction}

\renewcommand{\arraystretch}{1.4}

The list of vectorial Lie algebras (i.e. algebras of polynomial vector 
fields)
was conjectured by Lie and proven by Cartan \cite{Car09}. The analogous
problem in the super case was raised by Kac \cite{Kac77} and completed
in \cite{Kac98}; see also 
\cite{ALS80,BL81,CK97,CK99,Kac97,Lar01b,Lei85,LSh88,Lei90,LSh00,Sh83,Sh97,Sh99,ShP98}.
The list consists of ten infinite series and five exceptions.
In the present paper, I consider the three exceptional algebras
$\ksle(5|10)$, $\vle(3|6)$ and $\mb(3|8)$, where the numbers
indicate the super-dimensions of the spaces on which the algebras are
realized. Kac and collaborators denote these algebras $E(5|10)$,
$E(3|6)$ and $E(3|8)$. I use the names designed by Shchepochkina, who
first found these algebras, albeit implicitly and in inconsistently 
regraded form ($\vle(4|3)$ and $\mb(4|5)$) \cite{Sh83}. 
Kac \cite{Kac98} and Cheng and Kac \cite{CK99} gave the descriptions
as abstract Lie superalgebras (i.e. generators and brackets were written 
down), and the realizations as concrete subalgebras of 
$\vect(\nm)$ ($\nm = 5|10$, $3|6$ and $3|8$) preserving certain equations
first appeared in \cite{Lar01b}.

Every vectorial algebra $\oj$ admits a grading of depth $d$ and height 
$h$:
\[
\oj = \oj_{-d} + ... +\oj_{-1} + \oj_0 + \oj_1 + ... + \oj_h.
\]
In particular, any finite-dimensional Lie superalgebra with 
finite-dimensional representations is a vectorial algebra of depth and
height $0$, because every such algebra may be embedded into $gl(\nm)$, 
which admits the realization $x^\mu\d_\nu$. It has recently been 
observed \cite{GKN01} that all simple Lie algebras have a regrading 
of depth and height $1$ (``conformal realization''), except for $E_8$, 
$F_4$ and $G_2$, which instead have a regrading of depth and height 
$2$ (``quasiconformal realization''). In fact, every simple 
finite-dimensional Lie superalgebra have regradings of depth and
height at most $2$.
Infinite-dimensional vectorial algebras have infinite height and finite
depth. Apart from an inconsistent regrading of $\ksle(5|10)$, $\mb(3|8)$
is the unique simple Lie superalgebra of maximal depth $3$ \cite{Kac98}.
The grading is said to be {\em consistent} if the odd subspaces are purely 
fermionic and the even subspaces purely bosonic. It is known that the only
consistently graded simple algebras are the contact algebras $\kk(1|m)$
(a.k.a. the centerless $N=m$ superconformal algebra),
the three exceptions considered in the present paper, and a fourth 
exception $\kas(1|6)$ \cite{Kac98,LSh00}.

The algebras under consideration here can be described as Cartan 
prolongs. This means that one fixes a realization of the 
finite-dimensional algebra $\oj_0$ and the nilpotent algebra 
$\oj_- = \oj_{-d} + ... +\oj_{-1}$
(which is also a $\oj_0$ module) as vector fields acting on $\CC^\nm$,
and define $\oj_k$ recursively for positive $k$ as the maximal subspace of 
$\vect(\nm)$ satisfying $[\oj_k, \oj_{-1}] \subset \oj_{k-1}$. 
The prolong is denoted by 
$\oj = (\oj_{-d}, ..., \oj_{-1}, \oj_0)_* \equiv (\oj_-, \oj_0)_*$. 
We have
\[
\barr{l}
\vle(3|6)
 = ((\brep 3,\brep1,-2), (\brep3^*,\brep2,-1), \ssg)_*, \\
\ksle(5|10)
 = ((\brep 5,-2), (\brep{10}^*,-1), sl(5))_*, \\
\mb(3|8) 
 = ((\brep1,\brep2,-3), (\brep 3,\brep1,-2), 
 (\brep 3^*,\brep 2,-1), \ssg)_*. \\
\earr
\]
Here $\brep n$ denotes the $n$-dimensional representation of $sl(n)$,
$\brep n^*$ its dual, and $\brep{10}^* = \brep5^*\wwedge\brep5^*$ 
is the ten-dimensional $sl(5)$ module. The spaces $\oj_{-k}$ are described
as $\oj_0$ modules.

The explicit description of the algebras as subalgebras of vector fields
is based on the following two observations:
\begin{enumerate}
\item
Vector fields which preserve some structure, be it a differential form,
a fixed vector field, or equations involving forms or vector fields 
(Pfaff equations), automatically generate a closed subalgebra of 
$\vect(\nm)$.
\item
The prolong $\oj = (\oj_-, \oj_0)_*$ is completely determined by $\oj_-$
and $\oj_0$.
\end{enumerate}
The idea in \cite{Lar01b} was then to redefine the Cartan prolong in the
following geometrical way:
\begin{enumerate}
\item
Find a realization of $\oj_-$ and $\oj_0$ in $\nm$-dimensional space.
\item
Find the maximal set of structures preserved by this realization.
\item
Define $\oj$ as the full subalgebra of $\vect(\nm)$ preserving those
structures.
\end{enumerate}

Since the prolong $\oj$ is completely determined by $\oj_0\ltimes\oj_-$,
it is not surprising that there is a 1-1 correspondence between 
irreducible $\oj_0$ and $\oj$ modules. This correspondence can be 
described as follows. Given a $\oj_0$ module $V$, one can construct the 
corresponding tensor module $T(V)$, which is equivalent to (more 
precisely: contragredient to) the induced module
$U(\oj)\otimes_{U(\oj_0)}V$. If $T(V)$ is irreducible, which is often 
the case, we are done. Otherwise, a morphism $\nabla: T(V) \to T(V')$
exists, and the module $\ker\nabla$ may be irreducible; if not, 
more morphisms must be found. Since cohomology is almost always absent,
we may rewrite $\ker\nabla = \im\nabla'$, where 
$\nabla': T(V'') \to T(V)$ is another morphism.

A well-known example is $\vect(n) = (\brep n, gl(n))_*$. Its tensor 
modules are tensor densities, which are irreducible except for totally 
anti-symmetric tensor fields of condegree zero, i.e. differential 
forms. In this case one morphism, the exterior derivative $d$, exists. 
The space $\ker d$ consists of closed forms and $\im d$ of exact forms; 
for polynomials there is no difference due to Poincar\'e's lemma. With 
this geometrical picture in mind, I refer to reducible tensor 
modules as {\em form modules}, and their irreducible quotients as 
{\em closed form modules}.

The contragredient problem of finding singular vectors in induced modules
was considered by Kac and Rudakov for $\vle(3|6)$ \cite{KR00a,KR00b},
and $\mb(3|8)$ and $\ksle(5|10)$ \cite{KR01}. In their notation,
induced, degenerate and irreducible degenerate modules are equivalent to 
tensor, form and closed form modules, respectively.
The same problem was also 
considered by Grozman, Leites and Shchepochkina \cite{GLS01}, but 
unfortunately I have not yet understood their paper.

In the present paper I address the same problem from the more geometrical 
viewpoint of tensor modules. For $\vle(3|6)$ and $\mb(3|8)$ I find the
same results as Kac and Rudakov (up to contragredience), but my result for
$\ksle(5|10)$ differs significantly from theirs. I claim that the 
morphisms in their C sector are not invariant, and also that there 
are two families of invariant operators apart from their A and B 
sectors. I also describe an additional morphism for all algebras: the 
integral, which even is $\vect(\nm)$ invariant. However, the integral is 
not local when $n\geq1$, i.e. it is not a differential operator.

In fact, I only partially prove my results; all first-order operators 
($\nabla_1$) are shown to be $\oj$ invariant, but the higher-order 
operators are only shown to be $\oj_0$ invariant. The latter is very easy
to prove in tensor formalism, because it only amounts to matching 
upper and lower indices. To see how this comes about, it is instructive 
to consider the simpler case $\vect(n)$. Let $x = (x^i)$, $i=1,2,...,n$
denote the coordinates in $\CC^n$, let $\d_i = \d/\d x^i$ be the 
corresponding derivative, and let $X=X^i(x)\d_i$ be a vector field. 
In particular, $\oj_0 = gl(n)$ is generated by vector fields of the form
$x^i\d_j$. The differential $\al^i = dx^i$, its dual $\bal_j$, and the 
volume form $v=dx^1 dx^2... dx^n$ transform as
\be
\LX \al^i = \d_j X^i \al^j, \qquad
\LX \bal_j = -\d_j X^i \bal_i, \qquad
\LX v = \d_i X^i v.
\ee
We now introduce two gradings: the {\em weight} $z$ and the 
{\em degree} $\deg$, defined as in the following table:
\[
\barr{l|ccccc}
&x^i & \d_i & \al^i & \bal_j & v \\
\hline
z & 0 & 0 & 1 & -1 & n \\
\deg & 1 & -1& 1 & -1 & n \\
\earr
\]
We observe that the {\em exterior derivative} $d = \al^i\d_i$ is
the unique $\oj_0$-invariant differential operator of degree zero 
which can be constructed using the differential $\al^i$ and the 
derivatives of degree $-1$. 
This is important, because it turns out that all $\oj$-invariant morphisms
have degree zero, at least for the algebras under consideration in
this paper. Thus, we can very easily write down all candidate morphisms,
and then check $\oj$ invariance by hand.

The exterior derivative defines a morphism
\bes
d: && \Om^p \to \Om^{p+1} \nle
&& \om_\ip(x) \al^{i_1}..\al^{i_p}  \mapsto 
\d_j\om_\ip(x) \al^{i_1}..\al^{i_p}\al^j.
\eens
It is straightforward to prove that $d$ commutes with $\LX$, i.e.
$\LX d\om = d(\LX\om)$ for every $\om \in \Om^p$. 
$d$ also acts on the space of dual forms or polyvector
fields $\Om_p$. The invariant pairing between $\om\in\Om^p$ and 
$\ups\in\Om_p$ is given by integration:
$(\ups,\om) = \int \ups \om = \int \d^nx\ \om_\ip(x)\ups^\ip(x)$.
In this case we can identify $\Om_p$ with $\Om^{n-p}$ by 
$\ups^\ip(x) = \eps^{\ip j_1..j_{n-p}} \om_{j_1..j_{n-p}}(x)$, but such an
identification is impossible in the presence of odd coordinates 
\cite{BL81}. The action of the exterior derivative becomes
\bes
d: && \Om_p \to \Om_{p-1} \nle
&& \ups^\ip(x) v \bal_{i_1}..\bal_{i_p} \mapsto 
\d_j\ups^{i_1..i_{p-1}j}(x) v \bal_{i_1}..\bal_{i_{p-1}}
\eens
Finally, the integral defines an invariant morphism $\int: \Om_0\to\Om^0$.
We can thus summarize the situation as a differential complex:
\[
\barr{cccccc}
\Om_n \rarroww d \ldots \rarroww d \Om_0 \rarroww \int
\Om^0 \rarroww d \ldots \rarroww d \Om^n.
\earr
\]

In Sections 2--4 the analogous results for the exceptional superalgebras
are presented. Let $w_1^*$ and $w_d^*$ be the weights of the $\oj_0$ 
modules $\oj_{-1}$ and $\oj_{-d}$, respectively. It turns out that 
reducible tensor modules are built from tensor products of differentials
\bes
\gm^a \in w_1^*	&\qquad& \al^i \in w_d \nl
\bgm_a \in w_1 	&\qquad& \bal_i \in w_d^*,
\eens
where $d$ is the depth. However, dual weights ($w_1$ and $w_1^*$,
$w_d$ and $w_d^*$) must not appear in the same form\footnote{This 
is in disagreement with Kac' and Rudakov's result \cite{KR01}, 
whose C sector involves dual
weights.}. In addition, there is a scalar form $v$,
analogous to the volume form in the $\vect(n)$ case.
The weight can immediately be read off: $w_k$ carries weight
$k$, $w_k^*$ carries weight $-k$, and by abuse of notation we denote 
the weight of the scalar $v$ by $v$, too.
There are thus four sectors of candidate form modules:
\[
\barr{|c|cc|cc|cc|}
\hline
\hbox{Sector} && \hbox{$\oj_0$ weight} && z && \hbox{Basis} \\
\hline
A && m w_d + n w_1^* + kv && m d - n + kv
&&\al^{i_1}..\al^{i_m} \gm^{a_1}..\gm^{a_n} v^k \\
B && m w_d + n w_1 + kv && m d + n + kv
&&\al^{i_1}..\al^{i_m} \bgm_{a_1}..\bgm_{a_n} v^k \\
C && m w_d^* + n w_1^* + kv && - m d - n + kv
&&\bal_{i_1}..\bal_{i_m} \gm^{a_1}..\gm^{a_n} v^k \\
D && m w_d^* + n w_1 + kv && - m d - n + kv
&&\bal_{i_1}..\bal_{i_m} \bgm_{a_1}..\bgm_{a_n} v^k.\\
\hline
\earr
\]
To establish that a candidate form module is indeed reducible, one 
must check explicitly
that a morphism intertwining the $\oj$ action exists. This typically
only happens for a specific value of $k$, the power of the scalar form,
and thus we obtain a two-parameter family of form modules.
When depth $d=1$, the situation degenerates because $w_1$ and $w_d$ are
identical, and there is thus only two sectors. For $\vect(n)$, the
form modules $\Om^p$ correspond to the D sector and $k=0$, and the
dual modules $\Om_p$ are given by the A sector with $k=1$.

For $\vle(3|6)$ and $\mb(3|8)$, the degree zero subalgebra $\oj_0 = \ssg$,
i.e. the non-compact form of the symmetries of the standard model in
particle physics. This suggests that these algebras may have important
applications to physics \cite{Kac99,Lar01a}\footnote{
Note that the definition of $\mb(3|8)$ in \cite{Lar01a} is flawed.}. 
To my knowledge, this is the only place where
the standard model algebra arises naturally and unambigously in a 
mathematically deep context. Note that $\oj_0$ is not just any subalgebra
of $\oj$, but that there is a 1-1 correspondence between $\oj$ and $\oj_0$
modules. Therefore, one may speculate that a $\oj$ symmetry may be 
mistaken experimentally for a $\oj_0$ symmetry. 

With this motivation, in Section 5 I attempt to construct a gauge 
theory based on $\oj$ rather than $\oj_0$.
Given the enormous experimental success of the standard model,
any such theory must be very similar to it. The main difference is that 
the fermionic fields in an ordinary gauge theory are functions $\psi(x)$ 
depending on the spacetime coordinate $x$, whereas the $\oj$-invariant
theory involves fields $\psi(x,y)$ depending also on the coordinates $y$ 
of an internal $3|6$- or $3|8$-dimensional supermanifold. The addition of
this internal space will be called {\em second gauging}, for the 
following reason.
Start with a rigid symmetry $\oj_0$. The usual (first) gauging replaces 
$\oj_0$ with the algebra $\map(N,\oj_0)$ of maps from $N$-dimensional
spacetime to $\oj_0$. This makes the symmetry local in spacetime, but it
is still rigid in internal space. To make the symmetry local in internal
space as well, we replace $\oj_0$ with a prolong $\oj=(\oj_-,\oj_0)_*$.
This second gauging is of course highly non-unique, since the prolong 
depends on the nilpotent algebra $\oj_-$ in addition to $\oj_0$. However, 
if we further require that $\oj$ be simple, there are only two 
possibilities for $\oj_0 = \ssg$: $\vle(3|6)$ and $\mb(3|8)$. 

The fermions $\psi(x,y)$, valued in $\oj_0$ modules, are simply the tensor 
fields of $\map(N,\oj)$. However, it does not make much sense to pass
from $\oj_0$ to $\oj$ if we were only to consider tensor fields, since all
information about these is already encoded in $\oj_0$. Therefore, we
assume that a $\oj$-invariant morphism $\nabla$ exists, i.e. we identify 
(at least some of) the fermionic fields with the closed form modules 
constructed in Sections 2 and 3. 
However, the na\"\i ve identification of closed form modules with quarks 
and leptons does not quite work out neither for $\vle(3|6)$ nor for 
$\mb(3|8)$. 
It is unclear to me how the quantum numbers of the fundamental fermions 
should be related to the closed forms, if it is at all possible.
Nevertheless, some experimental predictions are generic: absense of
extra gauge bosons beyond the twelve already present in the standard 
model, and particle/anti-particle asymmetry (CP violation). Despite being
based on superalgebras, second-gauged theories are presumably not 
plagued with supersymmetric partners.

Throughout this paper tensor calculus notation is used. $A^i$ denotes
a contravariant vector and $B_j$ a covariant vector. Repeated indices,
one up and one down, are implicitly summed over (Einstein convention). 
Derivatives are denoted by various types of d's ($d$, $\d$ and $\eth$).
$\eps^{ab}$, $\eps^{ijk}$ and $\eps^{ijklm}$ denote
the totally anti-symmetric constant symbols in $\CC^2$, $\CC^3$ and
$\CC^5$.
When dealing with the non-positive subalgebra $\oj_0\ltimes\oj_-$, all
parities are known and it is convenient to explicitly 
distinguish between anti-symmetric (straight) and symmetric (curly) 
brackets: $[A,B] = -[B,A]$ and $\{A,B\} = \{B,A\}$.
For general vector fields $X=X^i(x)\d_i$ (always assumed to be 
homogeneous in parity), the (straight) brackets are graded in the usual 
way: $[X,Y] = -(-)^{XY}[Y,X]$, where the symbol $(-)^X$ is $+1$ on 
bosonic components and $-1$ on fermionic ones. 
The sign convention is that $X$ acts as
\bes
\LX\d_j &=& -(-)^{j X}\d_j X^i\d_i, \nle
\LX dx^i &=& (-)^{(X+i+j)j} \d_j X^i dx^j.
\eens

There are some special relations valid in two dimensions only, which are
needed in Sections 2 and 3:
\bes
&&\phi^a = \eps^{ab}\phi_b, \qquad \phi_a = \eps_{ab}\phi^b, 
\label{raise}\\
&&\eps^{ab}\eps_{bc} = \dlt^a_c, \qquad 
\eps^{ab}\eps_{cb} = -\dlt^a_c, \\
&&\phi^a\psi_a = -\phi_a\psi^a, \\
&&\eps^{ab}\phi^c + \eps^{bc}\phi^a + \eps^{ca}\phi^b = 0,
\label{ident}\\
&&\phi^a\psi^b - \phi^b\psi^a = -\eps^{ab}\phi^c\psi_c.
\label{skew}
\ees
Here $\eps^{ab}$ is the totally skew constant tensor in two dimensions
and $\eps_{ab}$ its inverse;
$\eps^{12} = \eps_{21} = +1$ and $\eps^{21} = \eps_{12} = -1$. 
We use these constants to raise and lower $sl(2)$ indices.

\section{$\vle(3|6)$}

Consider $\CC^{3|6}$ with basis spanned by three even coordinates
$u^i$, $i=1,2,3$ and six odd coordinates $\th_{ia}$.
Let $\deg \th_{ia} = 1$ and $\deg u^i = 2$.
The graded Heisenberg algebra has the non-zero relations
\be
[\d_j, u^i] = \dlt^i_j, \qquad 
\{d^{ia}, \th_{jb}\} = \dlt^i_j\dlt^a_b,
\ee
where $\d_i = \d/\d u^i$ and $d^{ia} = \d/\d\th_{ia}$.
Consider the vector fields\footnote{In \cite{Lar01b} these vector fields
were denoted by $\tilde E_i$ and $\tilde D^{ia}$, whereas the untilded
notation was reserved for the generators of $\oj_-$. Since the latter
is of no particular interest in the present paper, the tildes on the 
former are dropped to avoid unnecessarily cumbersome notation.}
\be
E_i = \d_i, \qquad
D^{ia} = d^{ia} + \eps^{ijk}\th^a_j\d_k,
\ee
which generate the nilpotent algebra $\toj_-$:
\bes
\{D^{ia}, D^{jb}\} &=& 2\eps^{ijk}\eps^{ab}E_k, \nle
{[}D^{ia}, E_j] &=& [E_i, E_j] = 0.
\eens
Any vector field in $\vect(3|6)$ has the form
\be
X = Q^i\d_i + P_{ia}d^{ia} = \tQ^i\d_i + P_{ia}D^{ia},
\ee
where
\be
\tQ^i &=& Q^i - (-)^X\eps^{ijk}\th^a_jP_{ka}. 
\label{vleQ}
\ee
$\vle(3|6)$ is the subalgebra of $\vect(3|6)$ which preserves the dual 
Pfaff equation $D^{ia}=0$, i.e. $X\in\vle(3|6)$ iff
\be
[X, D^{ia}] = -(-)^X D^{ia}P_{jb}D^{jb}.
\label{vleXD}
\ee
This leads to the condition
\be
D^{ia}\tQ^j = -2(-)^X\eps^{ijk}P^a_k.
\label{vleeqn}
\ee
In particular, we have the symmetry relations
\be
D^{ia}\tQ^j = -D^{ja}\tQ^i.
\label{vlesymm}
\ee
Alternatively, $\vle(3|6)$ preserves the form
\be
\al^i = \du^i - \eps^{ijk}\th^a_j\dth_{ka}
\label{vlePfaff}
\ee
up to a factor, i.e.
\be
\LX\al^i = \d_j\tQ^i\al^j.
\label{vleXa}
\ee
Explicitly,
\be
X = V_f = f^i\d_i -\fourth(-)^f\eps_{ijk}D^i_af^jD^{ka},
\ee
where $f = f^i(\th,u)\d_i$ is a vector field acting on $\CC^{3|6}$ and 
$D^{ia}f^j = -D^{ja}f^i$.

The $\vle(3|6)$ tensor modules are labelled by the $\oj_0 = \ssg$ weights
$(p,q; r; z)$, $p,q,r\in \NN$, $z\in\ZZ$, where $p\pi_1 + q\pi_2$ is an 
$sl(3)$ weight, $r$ is an $sl(2)$ weight and $z$ is a $gl(1)$ weight (the
eigenvalue of the grading operator). 
A typical element in the tensor module $T(p,q; r; z)$ has the form
$f(\th,u)\phi^{\ip|\ar}_\jq$, where $f(\th,u)$ is a polynomial function and
$\phi^{\ip|\ar}_\jq$ is totally symmetric in $\ip$, $\jq$ and $\ar$.
{F}rom (\ref{vleXD}) and (\ref{vleXa}) we see that among tensor modules,
the differentials $\al^i \in T(1,0; 0; 2)$ and $\gm^{ia} \in T(1,0; 1; -1)$
transform particularly simply:
\be
\LX\al^i = \d_j\tQ^i\al^j, \qquad
\LX\gm^{ia} = -(-)^X D^{ia}P_{jb}\gm^{jb}.
\label{vlediff}
\ee
Dual differentials $\bal_i\in T(0,1; 0; -2)$ and $\bgm_{ia}\in T(0,1;1;1)$
transform as
\be
\LX\bal_i = -\d_i\tQ^j\bal_j, \qquad
\LX\bgm_{ia} = (-)^X D^{jb}P_{ia}\bgm_{jb}.
\label{vledual}
\ee
If we assume that $\al^i$ and $\bgm_{ia}$ are fermions, we can construct
a density form $v = \al^1\al^2\al^3 = 
\bgm_{11}\bgm_{12}\bgm_{21}\bgm_{22}\bgm_{31}\bgm_{32}$;
$v \in T(0,0;0;6)$ and its dual $\bv\in T(0,0;0;-6)$ transform as
\bes
\LX v &=& \d_i\tQ^iv = (-)^X D^{ia}P_{ia}v, 
\nlb{vlev}
\LX\bv &=& -\d_i\tQ^i\bv = -(-)^X D^{ia}P_{ia}\bv.
\eens
The weight $z$ and the degree $\deg$ are defined by
\[
\barr{l|cccccccccc}
&\th_{ia} & u^i & d^{ia} & \d_i & \al^i & \bal_i &\gm^{ia} & \bgm_{ia}
& v & \bv \\
\hline
z & 0 & 0 & 0 & 0 & 2 & -2 & -1 & 1 & 6 & -6 \\
\deg & 1 & 2 & -1 & -2 & 2 & -2 & -1 & 1 & 6 & -6 \\
\earr
\]
According to the principles set out in the introduction, candidate
$\oj$-invariant morphisms are differential operators which only involve 
the fermionic derivative $D^{ia}$, have degree zero, 
and are invariant under $\oj_0 = \ssg$. The complete list of such
morphisms is given by\footnote{
Kac and Rudakov \cite{KR00b} find two different morphisms 
$\nabla_4'$ and $\nabla_4''$, but they seem to be identical, although they
of course act on different spaces.}
\bes
\nabla_1 &=& \bgm_{ia} D^{ia}, \nl
\nabla_2 &=& \eps_{ab} \bal_i \bal_j v D^{ia} D^{jb}, \nl
\nabla_3 &=& \eps_{ijk}\bgm_{la} \bgm_{mb} \bgm_{nc} \al^l\al^m\al^n
 \bv D^{ia} D^{jb} D^{kc}, 
\nlb{vlemorph}
\nabla_4 &=& \eps_{ijk}\eps_{cd} \bgm_{la} \bgm_{mb} \al^m
 D^{ia} D^{jb} D^{kc} D^{ld}, \nl
\nabla_6 &=& \eps_{ijk} \eps_{lmn} \eps_{be}\eps_{cf} \bgm_{pa}\bgm_{qd}
\al^p\al^q D^{ia} D^{jb} D^{kc} D^{ld} D^{me} D^{nf},\nl
\intth &=& \eps_{ijk} \eps_{lmn} \eps_{ad}\eps_{be}\eps_{cf}v 
 D^{ia} D^{jb} D^{kc} D^{ld} D^{me} D^{nf}
\eens
It is straightforward to check that any other candidate morphism is 
either identically zero or the composite of two morphisms from the 
list (\ref{vlemorph}).
The weights of the candidate morphisms can immediately be read off;
it is also indicated in the subscript.
\be
\barr{c|cccccc}
&\nabla_1 & \nabla_2 & \nabla_3 & \nabla_4 &\nabla_6 & \intth \\
\hline
z & 1 & 2 & 3 & 4 & 6 & 6
\earr
\label{vledeg}
\ee

Form modules can be constructed from the differentials (\ref{vlediff}),
(\ref{vledual}) and (\ref{vlev}).
The action of the morphisms on the differentials and their duals is given
by
\be
\al^i \bal_j = \bal_j\al^i = \dlt^i_j, \qquad
\gm^{ia}\bgm_{jb} = \bgm_{jb}\gm^{ia} = \dlt^i_j\dlt^a_b, \qquad
v \bv = \bv v = 1,
\ee
i.e. $\bal_i = \d/\d\al^i$, $\bgm_{ia}=\d/\d\gm^{ia}$, $\bv = \d/\d v$, 
etc. Therefore, a form may not contain differentials and dual 
differentials of the same type. The basic form is
\bes
\om &=& \om^{\iq|\ar}(\th,u)
\bal_{i_{r+1}}...\bal_{i_q}\bgm_{i_1a_1}...\bgm_{i_ra_r}, 
\nlb{om_D}
&=& \om^{\iq|\ar}(\th,u)
\al^{i_{q+1}}...\al^{i_r}\bgm_{i_1a_1}...\bgm_{i_ra_r},
\eens
where $\om^{\iq|\ar}(\th,u)$ is a polynomial function, completely symmetric
in $\iq$ and $\ar$, and the two 
expressions hold when $q\geq r$ and $q\leq r$, respectively.
Clearly, $\om\in T(0,q; r; -2q+3r) \equiv \Om_D(q,r)$.
For generic $q,r$, we assume that the morphism is first order in the 
fermionic derivative $D^{ia}$. Inspection of the list (\ref{vlemorph})
shows that it must be of the form $\nabla_1:\Om_D(q,r)\to\Om_D(q+1,r+1)$.
Explicitly,
\be
\nabla_1\om = \symm(D^{jb}\om^{i_1..i_q|a_1...a_r})
\bal_{i_{r+1}}...\bal_{i_q}\bgm_{i_1a_1}...\bgm_{i_ra_r}\bgm_{jb},
\label{nablaD}
\ee
if $q\geq r$, and a similar expression when $q\leq r$. 
$\symm()$ indicates that all indices inside the parentheses are 
symmetrized, e.g. 
\bes
\symm(D^{jb}\om) &=& D^{jb}\om, \nl
\symm(D^{jb}\om^{i|}) &=& (D^{jb}\om^{i|} + D^{ib}\om^{j|}), \nle
\symm(D^{jb}\om^{|a}) &=& (D^{jb}\om^{|a} + D^{ja}\om^{|b}), \nl
\symm(D^{jb}\om^{i|a}) &=& (D^{jb}\om^{i|a} + D^{ib}\om^{j|a} +
D^{ja}\om^{i|b} + D^{ia}\om^{j|b}).
\eens

Let us prove that this formula defines a morphism in the case $q=r=0$.
Then $\om = \om(\th,u)$ is an ordinary function (so $z=0$), and
$\nabla_1\om = D^{jb}\om\bgm_{jb}$. Now,
\be
\LX \nabla_1\om = (-)^X D^{jb}P_{ia}\bgm_{jb}D^{ia}\om
+ \bgm_{ia}X(D^{ia}\om),
\ee
and
\bes
(-)^X\nabla_1(\LX\om) &=& (-)^X\bgm_{jb}D^{jb}(X\om) \nle
&=& (-)^X\bgm_{jb}\Big( (-)^X XD^{jb}\om + (D^{jb}X)\om\Big).
\eens
It is straightforward to see that these expressions are equal,
using the definition (\ref{vleXD}) of $\vle(3|6)$ vector fields in the 
form $(D^{jb}X) = D^{jb}P_{ia}D^{ia}$. Hence 
$\LX\nabla_1 = (-)^X\nabla_1\LX$, i.e. $\nabla_1:\Om_D(0,0)\to\Om_D(1,1)$
is an $\vle(3|6)$-invariant morphism.
The proof that (\ref{nablaD}) is invariant is similar for non-zero $q$
and $r$. When $q\neq r$ one must use (\ref{vleXD}) and 
\be
D^{ia}E_k\tQ^j + D^{ja}E_k\tQ^i = E_k(D^{ia}\tQ^j + D^{ja}\tQ^i) = 0,
\label{vleDEQ}
\ee
in view of (\ref{vlesymm}) and $[D^{ia},E_k] = 0$.

Similarly, consider forms of the type
\bes
\om &=& \om_{\ip|\ar}(\th,u)
\al^{i_{r+1}}...\al^{i_p}\gm^{i_1a_1}...\gm^{i_ra_r}, \nle
&=& \om_{\ip|\ar}(\th,u)
\bal_{i_{p+1}}...\bal_{i_r}\gm^{i_1a_1}...\gm^{i_ra_r},
\eens
where $\om_{\ip|\ar}(\th,u)$ is a polynomial function and the two 
expressions hold when $p\geq r$ and $p\leq r$, respectively.
Clearly, $\om\in T(p,0; r; 2p-3r) \equiv \Om_A(p,r)$.
There is a morphism $\nabla_1: \Om_A(p,r)\to\Om_A(p-1,r-1)$, defined by
\be
\nabla_1\om = D^{jb}\om_{i_1..i_{p-1}j|a_1...a_{r-1}b}(\th,u)
\al^{i_r}...\al^{i_{p-1}}\gm^{i_1a_1}...\gm^{i_{r-1}a_{r-1}}
\label{nablaA}
\ee
when $p\geq r$, and a similar expression when $p\leq r$. 
It is straightforward to prove $\vle(3|6)$ invariance directly as in 
the $\Omega_D$ case, but it is easier to note that if $\om\in\Om_A(p,r)$
and $\ups\in\Om_D(p,r)$, there is an invariant pairing
\be
(\om,\ups) = \int \duth \om_{\ip|\ar}(\th,u)\ups^{\ip|\ar}(\th,u).
\ee
The pairing is invariant because the volume element has weight zero.
The morphism (\ref{nablaA}) can now be defined in terms of (\ref{nablaD})
by $(\nabla_1\om, \ups) = (\om, \nabla_1\ups)$.

There are two more types of form modules.
$\om \in \Om_B(p,r) \equiv T(p,0;r;2p+3r+6)$ has the form
\be
\om = \om_\ip^\ar(\th,u) \al^{i_1}...\al^{i_{p+r}}
\bgm_{i_{p+1}a_1}...\bgm_{i_{p+r}a_r}v
\ee
The morphism $\nabla_1: \Om_B(p,r)\to\Om_B(p-1,r+1)$ is given by
\be
\nabla_1\om = \symm(D^{jb} \om_{i_1..i_{p-1}j}^\ar(\th,u))
 \al^{i_1}...\al^{i_{p+r}}
 \bgm_{i_pa_1}...\bgm_{i_{p+r-1}a_r}\bgm_{i_{p+r}b} v.
\label{nablaB}
\ee
Note the presence of the scalar form $v$ which contributes $+6$ to
$z$; it is needed to make the morphism invariant.
Dually,	$\om \in \Om_C(q,r) \equiv T(0,q;r;-2q-3r-6)$ has the form
\be
\om = \om^\iq_\ar(\th,u) \bal_{i_1}...\bal_{i_{q+r}}
\gm^{i_{q+1}a_1}...\gm^{i_{q+r}a_r}\bv.
\ee
The morphism $\nabla_1: \Om_C(q,r)\to\Om_C(q+1,r-1)$ is given by
\be
\nabla_1\om = \symm(D^{jb} \om^\iq_{a_1..a_{r-1}b}(\th,u)) 
\bal_{i_1}...\bal_{i_{q+r-1}}\bal_j
\gm^{i_{q+1}a_1}...\gm^{i_{q+r-1}a_{r-1}}\bv.
\label{nablaC}
\ee

The form modules are summarized in the following table, where $p,q,r\geq0$
and the weight $z$ is the eigenvalue of the grading operator.
\be
\barr{|c|c|c|c|}
\hline
\hbox{Form}&z &\hbox{Basis} & \hbox{Condition}\\
\hline
\Om_A(p,r) & 2p-3r 
& \al^{i_{r+1}}...\al^{i_p}\gm^{i_1a_1}...\gm^{i_ra_r} & p\geq r\\
&& \bal_{i_{p+1}}...\bal_{i_r}\gm^{i_1a_1}...\gm^{i_ra_r}  & p\leq r\\
\Om_B(p,r) & 2p+3r+6
& \al^{i_1}...\al^{i_{p+r}}\bgm_{i_1a_1}...\bgm_{i_ra_r}v &\\
\Om_C(q,r) & -2q-3r-6
& \bal_{i_1}...\bal_{i_{q+r}}\gm^{i_1a_1}...\gm^{i_ra_r}\bv &\\
\Om_D(q,r) & -2q+3r 
& \bal_{i_{r+1}}...\bal_{i_q}\bgm_{i_1a_1}...\bgm_{i_ra_r} & q\geq r\\
&& \al^{i_{q+1}}...\al^{i_r}\bgm_{i_1a_1}...\bgm_{i_ra_r}  & q\leq r\\
\hline
\earr
\label{vleforms}
\ee

\begin{figure}
\begin{center}
\includegraphics{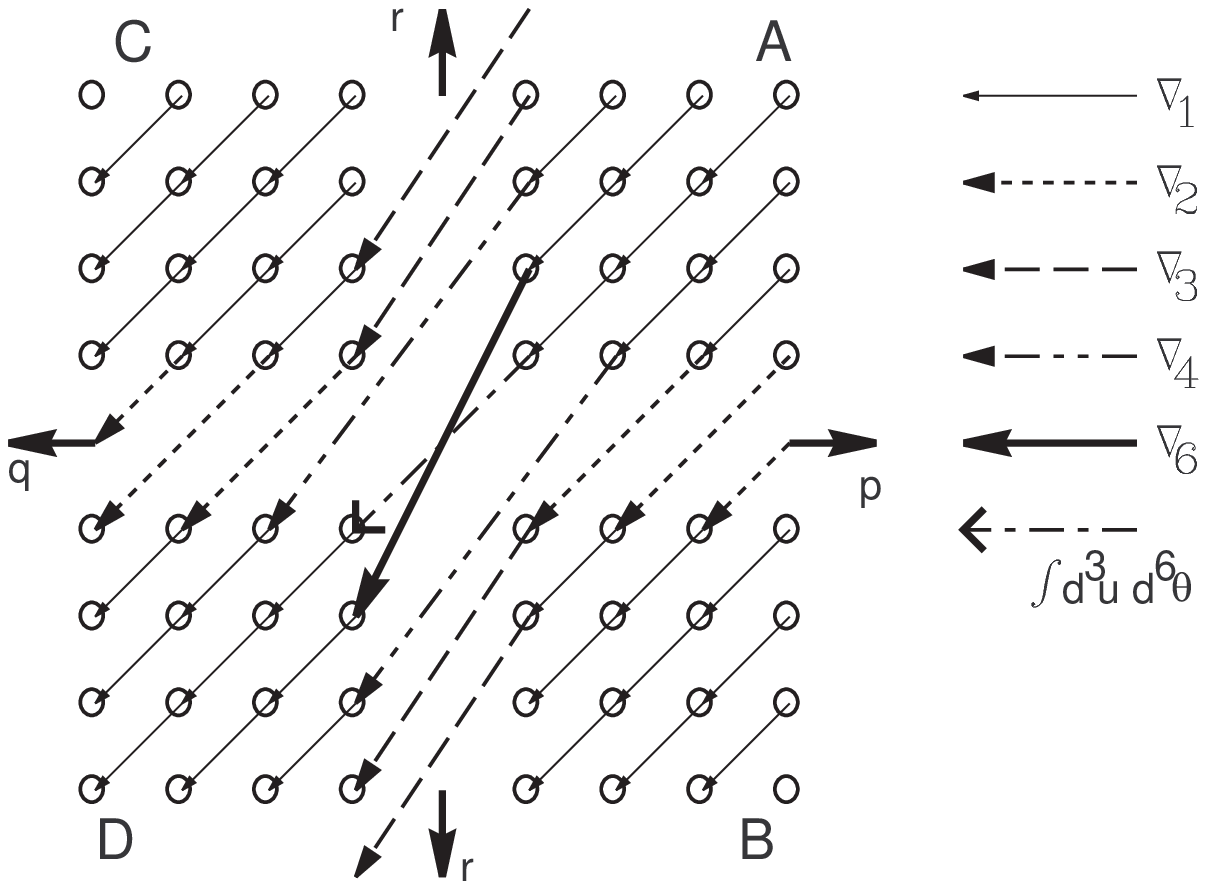}
  \caption{$\vle(3|6)$ form modules and morphisms. \hfill\break
  First quadrant: $\Om_A(p,r) = (p,r) = T(p,0; r; 2p-3r)$. \hfill\break
  Second quadrant: $\Om_B(p,r) = (p,-r) = T(p,0; r; 2p+3r+6)$. \hfill\break
  Fourth quadrant: $\Om_C(q,r) = (-q,r) = T(0,q; r; -2q-3r-6)$.\hfill\break
  Third quadrant: $\Om_D(q,r) = (-q,-r) = T(0,q; r; -2q+3r)$.
}
    \label{fig:vle}
  \end{center}
\end{figure}

The morphisms $\nabla_1$ acting on $\Om_A(p,r)$, $\Om_B(p,r)$ and 
$\Om_C(q,r)$ are only defined provided that $p\geq0$ and $r\geq0$ 
in the target module (this condition is automatically satisfied for
$\Om_D(q,r)$). Since all morphisms must be connected into infinite 
complexes (remember that the irreducible quotients can be written as
$\ker\nabla = \im\nabla'$), there must be some other morphisms acting on 
the remaining modules. By $\oj_0$ invariance, these morphisms must 
be taken from the candidate list (\ref{vlemorph}). 
All we have to do now is to fill the missing morphisms from the candidate
list, making sure to match all weights, including the $gl(1)$ weight $z$
given in (\ref{vledeg}). In particular, the factors of $v$ and $\bv$ must 
match. We find the following morphisms:
\be
\barr{rll}
\nabla_1: & \Om_A(p,r) \to \Om_A(p-1,r-1), & p\neq0, r\neq0\\
& \Om_B(p,r) \to \Om_B(p-1,r+1), & p\neq0\\
& \Om_C(q,r) \to \Om_C(q+1,r-1), & r\neq0\\
& \Om_D(q,r) \to \Om_D(q+1,r+1), &\\
\nabla_2: & \Om_A(p,0) \to \Om_B(p-2,0), & p\neq 0,1\\
& \Om_C(q,0) \to \Om_D(q+2,0), &\\
\nabla_3: & \Om_A(0,r) \to \Om_C(0,r-3), & r\neq 0,1,2\\
& \Om_B(0,r) \to \Om_D(0,r+3), & \\
\nabla_4: & \Om_A(0,2) \to \Om_D(1,0), & \\
& \Om_A(1,0) \to \Om_D(0,2), & \\
\nabla_6: & \Om_A(0,1) \to \Om_D(0,1), & \\
\earr
\label{vlenabla}
\ee
In addition, the integral defines a morphism
\bes
\int\duth:&& \Om_A(0,0) \to \Om_D(0,0), \nle
&& \om(\th,u) \to \int \duth \om(\th,u).
\eens
The integral is invariant under all of $\vect(3|6)$ and thus in particular
under $\vle(3|6)$. However, integration over bosonic coordinates is not 
a differential operator, which is why it was not seen by Kac and Rudakov
\cite{KR00a,KR00b}. Otherwise, my result is in perfect agreement with
theirs. The situation is summarized in Figure \ref{fig:vle}.

\section{$\mb(3|8)$}
Consider $\CC^{3|8}$ with basis spanned by three even coordinates
$u^i$, $i=1,2,3$, six odd coordinates $\th_{ia}$, and two
more odd coordinates $\vth^a$.
Let $\deg \th_{ia} = 1$, $\deg u^i = 2$ and $\deg \vth^a = 3$.
The graded Heisenberg algebra has the non-zero relations
\be
\{\eth_b, \vth^a\} = \dlt^a_b, \qquad
[\d_j, u^i] = \dlt^i_j, \qquad 
\{d^{ia}, \th_{jb}\} = \dlt^i_j\dlt^a_b,
\ee
where $\d_i = \d/\d u^i$, $d^{ia} = \d/\d\th_{ia}$, and 
$\eth_a = \d/\d\vth_a$.
The vector fields
\bes
D^{ia} &=& d^{ia} - 3\eps^{ijk}\th^a_j\d_k
+ \eps^{ijk}\th^a_j\th^b_k\eth_b - u^i\eth^a, \nl
E_i &=& \d_i - \th^a_i\eth_a, 
\label{mbDEF}\\
F^a &=& \eth_a,
\eens
satisfy the nilpotent Lie superalgebra $\toj_-$:
\bes
\{D^{ia}, D^{jb}\} &=& -6\eps^{ijk}\eps^{ab}E_k, 
\nlb{mb-}
{[}D^{ia}, E_j] &=& 2\dlt^i_j F^a, 
\eens
and all other brackets vanish.
Any vector field in $\vect(3|8)$ has the form 
\be
X = R^a\eth_a + Q^i\d_i + P_{ia}d^{ia}
= \tR^a\eth_a + \tQ^iE_i + P_{ia}D^{ia},
\label{mbRQP}
\ee
where
\bes
\tQ^i &=& Q^i + 3(-)^X\eps^{ijk}\th^a_jP_{ka}, 
\nlb{mbQR}
\tR^a &=& R^a + (-)^X\th^a_iQ^i 
+ 2\eps^{ijk}\th^a_i\th^b_jP_{kb} - u^iP^a_i.
\eens
$\mb(3|8)$ is the subalgebra of $\vect(3|8)$ which preserves the dual Pfaff
equation $D^{ia}=0$, where $D^{ia}$ was defined in (\ref{mbDEF}).
In other words, $X\in\mb(3|8)$ iff 
\be
[X, D^{ia}] = -(-)^X D^{ia}P_{jb}D^{jb},
\label{mbXD}
\ee
which leads to the conditions
\bes
D^{ia}\tQ^j &=& 6(-)^X\eps^{ijk}P^a_k,
\nlb{mbeqn}
D^{ia}\tR^b &=& -2(-)^X\eps^{ab}\tQ^i.
\eens
In particular, we have the symmetry relations
\bes
D^{ia}\tQ^j &=& -D^{ja}\tQ^i,
\label{mbsymm}\\
D^{ia}\tR^b &=& -D^{ib}\tR^a.
\eens
Alternatively, $\mb(3|8)$ may be defined as the subalgebra of $\vect(3|8)$
under which the forms
\bes
\al^i &=& \du^i + 3\eps^{ijk}\th^a_j\dth_{ka}, 
\nlb{mbPfaff}
\bt^a &=& \dvth^a - u^i\dth^a_i + \th^a_i\du^i 
+ 2\eps^{ijk}\th^a_i\th^b_j\dth_{kb}.
\eens
transform as
\bes
\LX\al^i &=& E_j\tQ^i\al^j - (-)^X\eth_a\tQ^i\bt^a, 
\nlb{mbXab}
\LX\bt^a &=& (-)^X\eth_b\tR^a\bt^b.
\eens
Explicitly,
\be
X &\equiv& M_f = f^a\eth_a +\fourth(-)^fD^{ia}f_aE_i
+ \hbox{${1\/48}$}\eps_{ijk}D^i_aD^{jb}f_bD^{ka},
\label{mbX}
\ee
where $f = f^a(\th,u,\vth)\eth_a$ is a vector field acting on 
$\CC^{3|8}$,  
$D^{ia}f^b = -D^{ib}f^a$ and
$D^{ia}D^{jb}f_b = - D^{ja}D^{ib}f_b$.
Note that $(-)^f = +1$ if $f$ is an even vector field, 
i.e. $f^a$ is an odd function.

The $\mb(3|8)$ tensor modules are labelled by the $\oj_0 = \ssg$ weights
$(p,q; r; z)$, $p,q,r\in \NN$, $z\in\ZZ$, where $p\pi_1 + q\pi_2$ is an 
$sl(3)$ weight, $r$ is an $sl(2)$ weight and $z$ is a
$gl(1)$ weight (the eigenvalue of the grading operator). 
A typical element in the tensor module $T(p,q; r; z)$ has the form
$f(\th,u)\phi^{\ip|\ar}_\jq$, where $f(\th,u)$ is a polynomial function 
and $\phi^{\ip|\ar}_\jq$ is totally symmetric in $\ip$, $\jq$ and $\ar$.
{F}rom (\ref{mbXD}) and (\ref{mbXab}) we see that among tensor modules,
the differentials $\al^i \in T(1,0; 0; 2)$, $\bt^a \in T(0,0; 1; 3)$
and $\gm^{ia} \in T(1,0; 1; -1)$ transform particularly simply:
\bes
\LX\al^i = E_j\tQ^i\al^j, \nl
\LX\bt^a = (-)^X\eth_b\tR^a\bt^b, 
\label{mbdiff}\\
\LX\gm^{ia} = -(-)^X D^{ia}P_{jb}\gm^{jb}.
\eens
The transformation law for $\al^i$ has been modified compared
to (\ref{mbXab}); this is permissible because $\bt^a$ transforms
irreducibly by itself.

Dual differentials $\bal_i\in T(0,1; 0; -2)$, $\bbt_a \in T(0,0; 1; -3)$
and $\bgm_{ia}\in T(0,1;1;1)$ transform as
\bes
\LX\bal_i = -E_i\tQ^j\bal_j, \nl
\LX\bbt_a = -(-)^X\eth_a\tR^b\bbt_b, \\
\LX\bgm_{ia} = (-)^X D^{jb}P_{ia}\bgm_{jb}.
\eens
If we assume that $\al^i$, $\bt^a$ and $\bgm_{ia}$ are fermions, we can 
construct a scalar density $v = \al^1\al^2\al^3 = \bt^1\bt^2 =
\bgm_{11}\bgm_{12}\bgm_{21}\bgm_{22}\bgm_{31}\bgm_{32}$;
$v \in T(0,0;0;6)$ and its dual $\bv\in T(0,0;0;-6)$ transform as
\bes
\LX v &=& E_i\tQ^iv = (-)^X\eth_a\tR^av = (-)^X D^{ia}P_{ia}v, \nle
\LX\bv &=& -E_i\tQ^i\bv = -(-)^X\eth_a\tR^a\bv = -(-)^X D^{ia}P_{ia}\bv.
\eens
Define the weight $z$ and the degree $\deg$ by
\[
\barr{l|cccccccccccccc}
&\th_{ia} & u^i & \vth^a & d^{ia} & \d_i & \eth_a 
& \al^i & \bal_j &\bt^a &\bbt_a&\gm^{ia} & \bgm_{ia} & v & \bv \\
\hline
z & 0 & 0 & 0 & 0 & 0 & 0& 2 & -2 & 3 & -3 & -1 & 1 & 6 & -6 \\
\deg & 1 & 2 & 3 & -1 & -2 & -3 & 2 & -2 & 3 & -3 & -1 & 1 & 6 & -6 \\
\earr
\]

As for $\vle(3|6)$, we build form modules using the elementary
differentials (\ref{mbdiff}).
The action of the morphisms on the differentials and their duals is given
by
\bes
\al^i \bal_j = \bal_j\al^i = \dlt^i_j, &\qquad&
\bt^a \bbt_b = \bbt_b \bt^a = \dlt^a_b, \nle
\gm^{ia}\bgm_{jb} = \bgm_{jb}\gm^{ia} = \dlt^i_j\dlt^a_b, &\qquad&
v \bv = \bv v = 1.
\eens
A form may not contain differentials and dual differentials of the
same type. Since there are three types of differentials, apart from the
scalar density $v$, one might
try to write down form depending on four parameters, e.g.
\be
\om = \om^{\iq|\ar}(\th,u,\vth)
\bal_{i_{s+1}}...\bal_{i_q}\bbt_{a_{s+1}}...\bbt_{a_r}
\bgm_{i_1a_1}...\bgm_{i_sa_s} v^k.
\ee
However, when one demands $\oj$ invariance it turns out that $\al^i$
and $\bal_j$ can not be used. Eq. (\ref{vleDEQ}) is replaced by
\be
D^{ia}E_k\tQ^j + D^{ja}E_k\tQ^i = 
2(\dlt^i_k F^a\tQ^j + \dlt^j_k F^a\tQ^i) \neq 0,
\label{mbDEQ}
\ee
because $[D^{ia}, E_j] = 2\dlt^i_j F^a$ from (\ref{mb-}).
Hence the candidate form modules are built only from differentials
$\bt^a$, $\gm^{ia}$, $v$, and their duals.

After tedious calculations, completely analogous to the $\vle(3|6)$
case, we find the following form modules:
\be
\barr{|c|c|c|c|}
\hline
\hbox{Form}&z &\hbox{Basis} & \hbox{Condition}\\
\hline
\Om_A(p,r) & -4p+3r-6 
& \gm^{i_1a_1}...\gm^{i_pa_p}\bt^{a_{p+1}}...\bt^{a_r}\bv & r\geq p\\
&& \gm^{i_1a_1}...\gm^{i_pa_p}\bbt_{a_{r+1}}...\bbt_{a_p}\bv & r\leq p\\
\Om_B(p,r) & -4p-3r-12
& \gm^{i_1a_{r+1}}...\gm^{i_pa_{r+p}}\bbt_{a_1}...\bbt_{a_{p+r}}\bv^2 &\\
\Om_C(q,r) & 4q+3r+6
& \bgm_{i_1a_{r+1}}...\bgm_{i_qa_{r+q}}\bt^{a_1}...\bt^{a_{q+r}}v &\\
\Om_D(q,r) & 4q-3r 
& \bgm_{i_1a_1}...\bgm_{i_qa_q}\bbt_{a_{q+1}}...\bbt_{a_r} & r\geq q\\
&& \bgm_{i_1a_1}...\bgm_{i_qa_q}\bt^{a_{r+1}}...\bt^{a_q} & r\leq q\\
\hline
\earr
\label{mbforms}
\ee
where $p,q,r\geq0$ and $z$ is the eigenvalue of the grading operator.

According to the principles set out in the introduction, candidate
$\oj$-invariant morphisms are differential operators which only involve 
the fermionic derivative $D^{ia}$, have degree zero, 
and are invariant under $\oj_0 = \ssg$. The complete list of such
morphisms is  obtained from (\ref{vlemorph}) by the substitutions
\be
\al^i = \gm^{ia}\bbt_a v, \qquad
\bal_i = \bgm_{ia}\bt^a \bv.
\ee
Thus,
\bes
\nabla_1 &=& \bgm_{ia} D^{ia}, \nl
\nabla_2 &=& \eps_{ab} \bgm_{ic}\bgm_{jd}\bt^c\bt^d \bv D^{ia} D^{jb}, \nl
\nabla_3 &=& \eps_{ijk}\bbt_a\bbt_b\bbt_c v^2 D^{ia} D^{jb} D^{kc}, 
\nlb{mbmorph}
\nabla_4 &=& \eps_{ijk}\eps_{cd} \bgm_{la} \bbt_b v
 D^{ia} D^{jb} D^{kc} D^{ld}, \nl
\nabla_6 &=& \eps_{ijk} \eps_{lmn} \eps_{be}\eps_{cf} \bbt_a\bbt_d v^2
D^{ia} D^{jb} D^{kc} D^{ld} D^{me} D^{nf},\nl
\intth &=& \eps_{ijk} \eps_{lmn} \eps_{ad}\eps_{be}\eps_{cf} v
 D^{ia} D^{jb} D^{kc} D^{ld} D^{me} D^{nf}.
\eens

To fill in the missing morphisms, we again choose from the candidate 
list above, and make sure to match powers of $v$. This works out in the
same way as for $\vle(3|6)$, except for the $\nabla_6$ arrow. 
The relevant weights are
\[
\barr{c|ccc}
&\nabla_6 & \Om_A(0,1) & \Om_D(0,1) \\
\hline
z(\vle(3|6)) & 6 & -3 & 3 \\
z(\mb(3|8)) & 6 & -3 & -3 \\
\earr
\]
Thus
\[
\nabla_6: \Om_A(0,1) \to \Om_D(0,1)
\]
is a well-defined $\vle(3|6)$ morphism, because 
$z(\nabla_6) + z_A(0,1) = z_D(0,1)$, but it is not an
$\mb(3|8)$ morphism. However, since $z_A(0,1) = z_D(0,1)$
in $\mb(3|8)$, the $\nabla_6$ arrow can be replaced by the identity map.

To summarize:

\begin{theorem}
The degenerate $\mb(3|8)$ tensor modules are
$\Om_A(p,r) =\softbreak T(p,0; r; -4p+3r-6)$, 
$\Om_B(p,r) = T(p,0; r; -4p-3r-12)$, 
$\Om_C(q,r) = T(0,q; r; 4q+3r+6)$, 
$\Om_D(q,r) = T(0,q; r; 4q-3r)$, ($p,q,r\geq0$).
The morphisms are the same as in (\ref{vlenabla}), except that 
the $\nabla_6$ arrow is replaced by the identity map (since 
$\Om_A(0,1) = \Om_D(0,1) = T(0,0; 1; -3)$).
The integral morphism reads
\bes
\int\duthv: \Om_A(0,0) &\to& \Om_D(0,0), \nl
\om(\th,u,\vth)\bv &\mapsto& \int \duthv \om(\th,u,\vth).
\eens
\end{theorem}

The morphisms can be illustrated by the Figure \ref{fig:vle}, which
is the same as for $\vle(3|6)$. This result is superficially different
from what Kac and Rudakov find \cite{KR01}, because they work in a 
contragredient formalism. Agreement is found if we redefine
$(p,q) \to (q,p)$, $z \to -z$, $\Om_A \leftrightarrow \Om_D$,
and $\Om_B \leftrightarrow \Om_C$\footnote{It is not clear to me why 
their arrows are not reversed.}. Again, they do not consider the
non-local integral, which explains the hole in the middle of their 
Figure 1.

\section{$\ksle(5|10)$}

Consider $\CC^{5|10}$ with basis spanned by five even coordinates
$u^i$, $i=1,2,3,4,5$ and ten odd coordinates $\th_{ij} = -\th_{ji}$.
Let $\deg \th_{ij} = 1$ and $\deg u^i = 2$.
The graded Heisenberg algebra has the non-zero relations
\be
[\d_j, u^i] = \dlt^i_j, \qquad \{d^{ij}, \th_{kl}\} 
= \dlt^i_k\dlt^j_l - \dlt^j_k\dlt^i_l,
\ee
where $\d_i = \d/\d u^i$ and $d^{ij} = -d^{ji} = \d/\d\th_{ij}$.
Consider the vector fields
\bes
D^{ij} &=& d^{ij} + \half\eps^{ijklm}\th_{kl}\d_m,
\nlb{ksleDE}
E_k &=& \d_k,
\eens
which generate the superalgebra
\bes
\{D^{ij}, D^{kl}\} &=& 2\eps^{ijklm}E_m, 
\nle
{[}D^{ij}, E_k] &=& [E_k, E_l] = 0.
\eens
Any vector field in $\vect(5|10)$ has the form
\be
X = Q^i\d_i + \half P_{ij}d^{ij} = \tQ^i\d_i + \half P_{ij}D^{ij},
\ee
where the $\half$ is necessary to avoid double counting and
\be
\tQ^i &=& Q^i + \mfourth\eps^{ijklm}\th_{jk}P_{lm}.
\label{ksleQ}
\ee

$\ksle(5|10)$ is the subalgebra of $\svect(5|10)$ (divergence-free vector
fields) which preserve the dual Pfaff equation $D^{ij}=0$. For 
convenience, we will adjoin the grading operator 
$Z=2u^i\d_i + \half\th_{ij}d^{ij}$ to $\ksle(5|10)$, and thus consider
the non-simple algebra $\bigksle = (\brep 5, \brep{10^*}, gl(5))_*$.
Hence $X\in\bigksle$ iff
\be
[X, D^{ij}] = -\mhalf D^{ij}P_{kl}D^{kl}
\label{ksleXD}
\ee
which leads to the condition
\be
D^{ij}\tQ^k = (-)^X\eps^{ijklm}P_{lm}.
\label{ksleeqn}
\ee
In particular, we have the symmetry relations
\be
D^{ij}\tQ^k = D^{jk}\tQ^i = -D^{ik}\tQ^j.
\label{kslesymm}
\ee
Alternatively, $\ksle(5|10)$ can be defined as the vector fields which
preserve the form
\be
\al^i = \du^i + \fourth\eps^{ijklm}\th_{jk}\dth_{lm}
\ee
up to a factor:
\be
\LX\al^i = \d_j\tQ^i\al^j.
\label{ksleXa}
\ee
Explicitly,
\be
X \equiv U_f = f^i\d_i + \hbox{$1\/24$}(-)^f\eps_{ijklm} D^{ij}f^kD^{lm},
\ee
where $f = f^i(\th,u)\d_i$ is a vector field acting on $\CC^{5|10}$,
$D^{ij}f^k = D^{jk}f^i = -D^{ik}f^j$, and $\d_if^i = 0$.

The $\bigksle$ tensor modules are labelled by the 
$sl(5)$ weights $(p,q,r,s) = p\pi_1 + q\pi_2 + r\pi_3 + s\pi_4$,
$p,q,r,s\in \NN$, 
and a $gl(1)$ weight $z\in\ZZ$ (the eigenvalue of the grading operator).
A typical element in the tensor module $T(p,q,r,s; z)$ has the form
$f(\th,u)\phi^{\ip|\klq}_{\js|\mnr}$, where $f(\th,u)$ is a polynomial 
function and $\phi^{\ip|\klq}_{\js|\mnr}$ is totally symmetric in 
$\ip$, $\klq$, $\mnr$, 
and $\js$, and anti-symmetric under $k_1l_1 \to l_1k_1$ and 
$m_1n_1 \to n_1m_1$, etc.
{F}rom (\ref{ksleXD}) and (\ref{ksleXa}) we see that among tensor modules,
the differentials $\al^i \in T(1,0,0,0; 2)$ and 
$\gm^{ij} \in T(0,1,0,0; -1)$ transform particularly simply:
\be
\LX\al^i = \d_j\tQ^i\al^j, \qquad
\LX\gm^{ij} = -\mhalf D^{ij}P_{kl}\gm^{kl}.
\label{kslediff}
\ee
Dual differentials $\bal_i\in T(0,0,0,1; -2)$ and 
$\bgm_{ij}\in T(0,0,1,0;1)$ transform as
\be
\LX\bal_i = -\d_i\tQ^j\bal_j, \qquad
\LX\bgm_{ij} = \mhalf D^{kl}P_{ij}\bgm_{kl}.
\ee
If we assume that $\al^i$ and $\bgm_{ia}$ are fermions, we can construct
a density form $v = \al^1\al^2\al^3\al^4\al^5 = 
\bgm_{12}\bgm_{13}\bgm_{14}\bgm_{15}\bgm_{23}
\bgm_{24}\bgm_{25}\bgm_{34}\bgm_{35}\bgm_{45}$;
$v \in T(0,0,0,0;10)$ and its dual $\bv\in T(0,0,0,0;-10)$ transform as
\bes
\LX v &=& \d_i\tQ^iv = \mhalf D^{ij}P_{ij}v, \nle
\LX\bv &=& -\d_i\tQ^i\bv = -\mhalf D^{ij}P_{ij}\bv.
\eens

Forms can be constructed from the differentials (\ref{kslediff}), with
the following action on the differentials and their duals:
\be
\al^i \bal_j = \bal_j\al^i = \dlt^i_j, \quad
\gm^{ij}\bgm_{kl} = \bgm_{kl}\gm^{ij} 
= \dlt^i_k\dlt^j_l - \dlt^i_l\dlt^j_k, \quad
v\bv = \bv v = 1.
\ee
Therefore, a form may not contain differentials and dual differentials 
of the same type. The basic form is
\bes
\om &=& \om^{\is|\jkr}(\th,u)
\bal_{i_1}...\bal_{i_s}\bgm_{j_1k_1}...\bgm_{j_rk_r}, 
\eens
where $\om_{\is|\jkr}(\th,u)$ is a polynomial function, totally symmetric
in $\is$, anti-symmetric under $j_mk_m\leftrightarrow k_mj_m$,
and symmetric under $j_mk_mj_nk_n \leftrightarrow j_nk_nj_mk_m$.
Clearly, $\om\in T(0,0,r,s; r-2s) \equiv \Om_D(s,r)$.
There is a morphism $\nabla_1: \Om_D(s,r)\to\Om_D(s,r+1)$, defined by
\be
\nabla_1\om = \symm(D^{lm}\om^{\is|\jkr}(\th,u))
\bal_{i_1}...\bal_{i_s}\bgm_{j_1k_1}...\bgm_{j_rk_r}\bgm_{lm},
\label{nablaksle}
\ee
where the indices $lm$ are symmetrized with $\jkr$, e.g.
\bes
\symm(D^{lm}\om^{\is|}) &=& D^{lm}\om^{\is|}, \nl
\symm(D^{lm}\om^{\is| j_1k_1}) &=& D^{lm}\om^{\is| j_1k_1}
+ D^{j_1k_1}\om^{\is| lm}, \nle
\symm(D^{lm}\om^{\is| j_1k_1j_2k_2}) &=& 
D^{lm}\om^{\is| j_1k_1j_2k_2} +\nl
&&+ D^{j_1k_1}\om^{\is| lmj_2k_2}
+ D^{j_2k_2}\om^{\is| lmj_1k_1}.
\eens
The verification that (\ref{nablaksle}) indeed defines a morphism
involves (\ref{ksleXD}) in the form
\be
(D^{ij}X) = \half D^{ij}P_{kl}D^{kl},
\ee
(\ref{kslesymm}) and $[D^{ij}, E_k] = 0$.

Using similar calculations one proves that there are four sectors
of form modules, as summarized in the following table. 
Here $p,q,r,s\geq0$ and the weight $z$ is the eigenvalue of the grading 
operator. 
\[
\barr{|c|c|c|c|}
\hline
\hbox{Form}&\hbox{$sl(5)$ weight}&z &\hbox{Basis} \\
\hline
\Om_A(p,q) & (p,q,0,0) & 2p-q 
& \al^{i_1}...\al^{i_p}\gm^{j_1k_1}...\gm^{j_qk_q}\\
\Om_B(p,r) & (p,0,r,0) & 2p+r 
& \al^{i_1}...\al^{i_p}\bgm_{j_1k_1}...\bgm_{j_rk_r}\\
\Om_C(s,q) & (0,q,0,s) & -2s-q 
& \bal_{i_s}...\bal_{i_s}\gm^{j_1k_1}...\gm^{j_qk_q}\\
\Om_D(s,r) & (0,0,r,s) & -2s+r 
& \bal_{i_1}...\bal_{i_s}\bgm_{j_1k_1}...\bgm_{j_rk_r}\\
\hline
\earr
\]
The first-order morphisms which have been constructed are of the form
\be
\barr{lll}
\nabla_1: & \Om_A(p,q) \to \Om_A(p,q-1), & q \neq 0 \\
 & \Om_B(p,r) \to \Om_B(p,r+1), & \\
 & \Om_C(s,q) \to \Om_C(s,q-1), & q \neq 0 \\
 & \Om_D(s,r) \to \Om_D(s,r+1), & \\
\earr
\label{ksle1}
\ee
Note that 
\bes
\Om_A(p,0) = \Om_B(p,0), &\qquad& \Om_A(0,q) = \Om_C(0,q),
\nlb{OmABCD}
\Om_D(s,0) = \Om_C(s,0), && \Om_D(0,r) = \Om_B(0,r),
\eens
because no explicit factors of the scalar differential $v$ appear.
Therefore, all morphisms are already connected into infinite complexes.
Nevertheless, there are more morphisms. First of all, the integral is
$\vect(5|10)$ invariant, so there is a map
\bes
\int \duthh: \Om_X(0,0) &\to& \Om_X(0,0), \nle
\om(\th,u) &\mapsto& \int \duthh \om(\th,u)
\eens
(the module at the origin is the same for all four sectors).
Moreover, there exist further differential operators that only involve
the fermionic derivative $D^{ij}$, have degree zero, 
and are invariant under $\oj_0 = sl(5)\ooplus gl(1)$. The complete list 
of such morphisms is given by
\bes
\nabla_1 &=& \bgm_{ij} D^{ij}, \nl
\nabla_2 &=& \eps_{ijklm}\eps_{npqrs}\al^i \al^n 
 D^{jp} D^{kq} D^{lr} D^{ms}, 
\label{kslemorph}\\
\intth &=& v D^{12}D^{13}D^{14}D^{15}D^{23}D^{24}D^{25}D^{34}D^{35}D^{45}.
\eens
It is natural to conjecture that $\nabla_2$ acts on the modules with
$q=0$ and $r=0$, which thus have two ingoing and two outgoing morphisms:
\[
\barr{rl}
\nabla_2: & \Om_A(p,0) \to \Om_B(p+2,0),  \\
\nabla_2: & \Om_C(s,0) \to \Om_D(s-2,0),  \\
\earr
\]

\begin{figure}
\begin{center}
\includegraphics{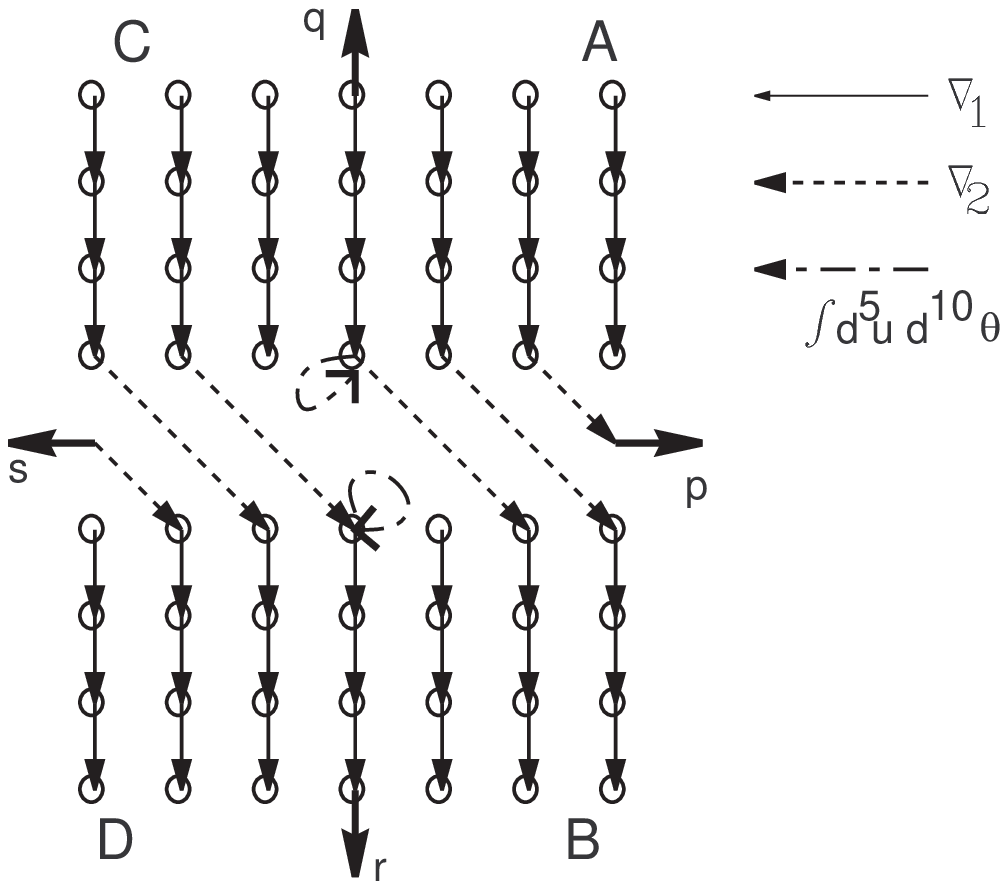}
  \caption{$\ksle(5|10)$ form modules and morphisms. The modules on the
  $q=0$ or $r=0$ axis has been split into two to give a clearer view of
  the arrows. The loop at the origin is the integral morphism. Existence
  of the arrows denoted by $\nabla_2$ is conjectural.
  \hfill\break
  First quadrant: $\Om_A(p,q) = (p,q) = T(p,q,0,0; 2p-q)$. \hfill\break
  Second quadrant: $\Om_B(p,r) = (p,-r) = T(p,0,r,0; 2p+r)$. \hfill\break
  Fourth quadrant: $\Om_C(s,q) = (-s,q) = T(0,q,0,s; -2s-q)$.\hfill\break
  Third quadrant: $\Om_D(s,r) = (-s,-r) = T(0,0,r,s; -2s+r)$.
}
    \label{fig:ksle}
  \end{center}
\end{figure}

The situation (conjectured for $\nabla_2$ but proven for $\nabla_1$)
is summarized in Figure \ref{fig:ksle}.

This result disagrees with what Kac and Rudakov find in \cite{KR01}. 
They find three sectors, with $sl(5)$ weights
\be
A = (p,q,0,0), B = (0,0,r,s), C = (p,0,0,s).
\ee
The first two sectors are the same as my $\Om_A$ and $\Om_D$ forms,
and the omission of $\Om_B$ and $\Om_C$ is not a problem, since
they only conjecture completeness of their list. However, I am not
able to verify that the tensor modules of type $(p,0,0,s)$ are 
reducible. The alleged morphism is, in the notation of the present paper,
\be
\nabla_C = (\bal_i{\d\/\d\al^j} - \bal_j{\d\/\d\al^i})D^{ij},
\ee
when acting on tensor fields of the form
\be
\om^\js_\ip \al^{i_1}...\al^{i_p} \bal_{j_1}...\bal_{j_s}.
\ee
When acting on pure powers, i.e. modules which involve either $\al^i$
or $\bal_j$ but not both, we have identified
$\al^i = \d/\d\bal_j$ and $\bal_j = \d/\d\al^i$. This identification is
not possible here, because it would make $\nabla_C \equiv 0$.
Moreover, I have verified that the simplest case, 
\bes
\nabla_C: (1,0,0,0) &\to& (0,0,0,1), \nle
 \om_i\al^i &\mapsto& D^{ij}\om_i\bal_j
\eens
is in fact not $\ksle(5|10)$ invariant.
Hence I conclude that tensor modules of type C are not form modules.

\section{ A second-gauged standard model }
The potential interest of the exceptional superalgebras is based on the 
following two observations:
\begin{enumerate}
\item
The grade zero subalgebra of $\vle(3|6)$ and $\mb(3|8)$ is $\oj_0 = \ssg$,
i.e. the non-compact form of the symmetries of the standard model in
particle physics. 
\item
$\mb(3|8)$ is the unique superalgebra of maximal depth $3$ in its 
consistent gradation. 
\end{enumerate}
Even if depth in the technical sense does not necessarily mean profound,
the technical and informal usages of the word are related; maximal
depth essentially means that the algebra has the most intricate and 
beautiful structure possible. 
The richest conceivable {\em local} symmetry is hence
intimately related to the symmetries of the standard model;
there is a 1-1 correspondence between irreducible modules.
It is therefore very tempting to construct a gauge theory with a local
$\mb(3|8)$ symmetry, and in this section a first step in this direction 
is taken. 
We do not limit ourselves to the exceptional superalgebras, but 
consider the general case of replacing a finite-dimensional symmetry 
$\oj_0$ by a quite general class of infinite-dimensional symmetries
$\oj$. Due to the shortage of letters, indices in this section are not
related to indices with identical names in previous sections.
Moreover, we ignore all superalgebra sign factors and pretend that we
are dealing with ordinary Lie algebras. This is of course not correct,
but keeping track of signs will only complicate matters and obscure 
the main points.

Let $J^a$, $a = 1, 2, ..., \dim \oj_0$ be the generators of the 
finite-dimensional Lie algebra 
$\oj_0\subset gl(\nm)$ with structure constants $f^{ab}{}_c$:
\be
[J^a, J^b] = f^{ab}{}_c J^c.
\label{JJ}
\ee
Let $y^i$, $i = 1, ..., n, n+1, ... n+m$ be the coordinates of $\CC^\nm$, 
and let $\d_i = \d/\d y^i$ be the corresponding derivatives. Then 
$\oj_0$ can be realized as vector fields of the form
$J^a = -\rep^{ja}_i y^i\d_j$.
Let $D_i = \d/\d y^i + ...$ be $n+m$ first order derivatives,
which generate a closed nilpotent algebra $\toj_-$ which is also a $\oj_0$
module:
\bes
[D_i, D_j] &=& g_{ij}{}^k D_k
\nlb{JD}
{[}J^a, D_j] &=& \rep^{ia}_j D_i.
\eens
The structure constants	$g_{ij}{}^k$ and $\rep^{ia}_j$ are assumed to 
respect the grading, i.e. $g_{ij}{}^k = 0$ unless 
$\deg y^i + \deg y^j = \deg y^k$
and $\rep^{ia}_j = 0$ unless $\deg y^i = \deg y^j$. 

Any vector field $X\in\vect(\nm)$ can be expanded in the $D_i$
basis: $X = X^i(y)D_i$. Clearly,
\be
[X,Y] = X^iD_iY^jD_j - Y^jD_jX^iD_i + g_{ij}{}^kX^iY^jD_k.
\label{XY}
\ee
Now consider the vector fields $X$ whose bracket with $D_i$ is of the 
form
\be
[D_i, X] \equiv D_iX^kD_k + g_{ij}{}^kX^jD_k
= \bar h^{jl}_{ik}D_jX^kD_l  
\label{Pfaff0}
\ee
for some structure constants $\bar h^{jl}_{ik}$ which also preserve the
grading: $\bar h^{jl}_{ik}=0$ unless $\deg y^j + \deg y^l = \deg y^i
+ \deg y^k$. Setting  $\bar h^{jl}_{ik} = \dlt^j_i\dlt^l_k +h^{jl}_{ik}$,
(\ref{Pfaff0}) can be rewritten as
\be
g_{ij}{}^lX^j = h^{jl}_{ik} D_jX^k.
\label{Pfaff}
\ee
This equation defines the subalgebra $\oj\subset\vect(\nm)$.

Define $\si^i_{ja}$ to be the inverse of the projection $\rep^{ia}_j$:
\be
\rep^{ia}_j\si^j_{ib} = \dlt^a_b, \qquad
\rep^{ia}_j\si^k_{la} \approx \dlt^i_l\dlt^k_j.
\ee
The notation $\approx$ is defined to mean equality when contracted with
$D_jX^i$. Thus, the second relation above should be interpreted to mean 
$D_jX^i = \rep^{ia}_j\si^k_{la}D_kX^l$ for every $X\in\oj$.
Note that this relation can not hold for arbitrary vector fields, but only
for vector fields in $\oj$.
It is now straightforward to prove that
\be
\LX = X + D_j X^i \si^j_{ia} T^a  
\equiv X + D_j X^i T^j_i
\label{LX}
\ee
satisfies $[\LX,\LY] = \LXY$, with $[X,Y]$ as in (\ref{XY}), provided that
\bes
\si^k_{ia}\si^l_{jb} f^{ab}{}_c &\approx&
\dlt^l_i\si^k_{jc} - \dlt^k_j\si^l_{ic} \nle
&& + h^{kl}_{mi}\si^m_{jc} - h^{lk}_{mj}\si^m_{ic}
+ h^{lm}_{ij}\si^k_{mc} - h^{km}_{ji}\si^l_{mc}
\eens                       
($\approx$ indicates that the relation holds when contracted with
$D_kX^iD_lY^j$, for all $X,Y\in\oj$.)
There are cases were relations of this type exist; the exceptional 
Lie superalgebras provide examples, which follows from the explicit 
descriptions of their tensor modules.
The generators $T^i_j$ satisfy $gl(\nm)$ weakly:
\be
[T^i_j, T^k_l] \approx \dlt^k_j T^i_l - \dlt^i_l T^k_j.
\ee
We thus have the projection
$\oj \onto gl(\nm): T^a = \rep^{ia}_j T^j_i$ and the injection
$\oj\into gl(\nm): T^i_j \approx \si^j_{ia}T^a$.
The transformation laws can be read off from (\ref{LX}); e.g., 
a vector and a covector transform as
\bes
\LX \gm^i &=& -X\gm^i + D_j X^i\gm^j, 
\nlb{gp1}
\LX \psi_j &=& -X\psi_j - D_j X^i\psi_i.
\eens

The symmetry of the gauge theory is not $\oj$ itself, but rather the
associated current algebra $\map(N,\oj)$; it does not matter 
if $\oj = \ssg$, $\vle(3|6)$, or $\mb(3|8)$ here. 
We must hence develop
the representation theory of the algebra of maps from $N$-dimensional
spacetime to an infinite-dimensional vectorial algebra $\oj$. There is
one important complication compared to the situation when $\oj$ is
finite-dimensional. 

Actually, we consider the larger algebra $\vectmapg$ obtained by adding
spacetime diffeomorphisms $\vect(N)$. This algebra is clearly a subalgebra
of $\vectmapv\subset\vect(N+\nm)$, the algebra of diffeomorphisms in
$N+\nm$-dimensional space which preserve the splitting into an 
$N$-dimensional base space and an $\nm$-dimensional fiber. Any such 
vector field can be decomposed into horizonal and vertical parts:
$(\xi,X) = (\xmu(x)\dmu, X^i(x,y)D_i)$,
where $x = x^\mu$, $\mu=1,2,...,N$, is the spacetime coordinate and
$\dmu = \d/\d x^\mu$ the corresponding derivative.
The bracket in $\vectmapg$ takes the form
\bes
[\xi,\eta] &=& \xmu\dmu\ynu\dnu - \ynu\dnu\xmu\dmu, \nl
{[}\xi, X] &=& \xmu\dmu X^iD_i, \\
{[}X,Y] &=& X^iD_iY^jD_j - Y^jD_jX^iD_i
+ g_{ij}{}^kX^iY^jD_k.
\eens
The corresponding Lie derivatives are
\bes
\Lxi &=& \xi + \dnu\xmu T^\nu_\mu, \nle
\LX &=& X + D_j X^i \si^j_{ia} T^a + \eps \dnu X^i T^\nu_i,
\eens
where 
\bes
[T^\mu_\nu, T^\rho_\si] &=& \dlt^\rho_\nu T^\mu_\si 
- \dlt^\mu_\si T^\rho_\nu, \nl
{[}T^\mu_\nu, T^\rho_j] &=& \dlt^\rho_\nu T^\mu_j, \nle
{[}T^a, T^\mu_j] &=& \rep^{ia}_j T^\mu_i, \nl
{[}T^\mu_i, T^\nu_j] &=& [T^\mu_\nu, T^a] = 0.
\eens
Note the appearence of the last term in the expression for $\LX$. The
constant $\eps$ can take on any value without violating the 
representation condition. The point is that if we set $\eps\neq0$, 
vectors in internal space acquire spacetime components as well.
In particular, (\ref{gp1}) is replaced by
\bes
\LX\gm^\mu &=& -X\gm^\mu, \nl
\LX \gm^i &=& -X\gm^i + D_j X^i\gm^j + \eps \dnu X^i\gm^\nu, 
\nlb{gp2}
\LX \psi_\nu &=& -X\psi_\nu - \eps\dnu X^i\psi_i, \nl
\LX \psi_j &=& -X\psi_j - D_j X^i\psi_i.
\eens
Thus $(\gm^\mu, \gm^i)$ and $(\psi_\nu,\psi_j)$ carry reducible but
indecomposable representations of $\map(N,\oj)$, with $\gm^\mu$ and 
$\psi_j$ being the irreducible submodules.
If we set $\eps=0$, assume that
$X$ is of the form $X = X_a(x)\rep^{ja}_iy^i\d_j$, and consider the
action on $y$-independent functions only, the expression for $\LX$
becomes $\LX = X_a(x)T^a$, which is recognized as the natural
realization of the current algebra $\map(N,\oj_0)$. However, when we
consider the full algebra $\oj$ the restriction to $y$-independent
functions is not meaningful.

The standard model enjoys extreme experimental success, so any attempt to
modify it must in some sense stay very close to it. A natural strategy
is to start from the standard model action and make the minimal 
modifications necessary to elevate the $\oj_0$ symmetry to a $\oj$
symmetry. Focus on the Yang-Mills-Dirac part:
\bes
S_D &=& \int \dNx \bpsi(x)\gm^\mu\DD_\mu\psi(x),
\nlb{S_D}
\DD_\mu &\equiv& \dmu + A_{a\mu}(x)T^a,
\eens
which leads to the massless Dirac equation $\gm^\mu\DD_\mu\psi(x) = 0$.
It is now tempting to simply replace the spacetime fields
with functions of both $x$ and $y$:
$\psi(x,y)$, $\bpsi(x,y)$ and $A_{a\mu}(x,y)$.
However, this does not work. One readily verifies that 
\bes
\LX \dnu\psi &=& -X\dnu\psi - D_jX^i\si^j_{ia}T^a\dnu\psi
\nlb{LXdp}
&&- \dnu X^iD_i\psi - \dnu D_jX^i\si^j_{ia}T^a\psi.
\eens
The last term can be compensated by demanding that the connection
$A_{a\mu}$ transforms as
\be
\LX A_{a\mu} = -XA_{a\mu} - f^{bc}{}_a X_bA_{c\mu}
- \dmu X^i A_{ai} + \si^j_{ia}\dmu D_jX^i,
\label{LXA}
\ee
but the third term forces us to introduce components $D_j\psi$ in the
internal directions as well. In other words, $(\dnu\psi, D_j\psi)$
transforms as the vector $(\psi_\nu, \psi_j)$ in (\ref{gp2}) with $\eps=1$.
There is no way that we can choose $\eps=0$ to make the vector's
spacetime components decouple from its internal components.

It is now easy to see that the simplest $\oj$-invariant generalization
of (\ref{S_D}) is
\bes
S_D &=& \int \dNx\dnmy \bpsi(x,y)
(\gm^\mu \DD_\mu + \gm^i(x,y)\DD_i)\psi(x,y), \nl
\DD_\mu &=& \dmu + A_{a\mu}(x,y)T^a,
\label{S_D2} \\
\DD_i &=& D_i + A_{ai}(x,y)T^a,
\eens
where $\LX$ acts on $\gm^\mu$ as (\ref{gp2}) with $\eps=1$ and on
$A_{a\mu}$ as (\ref{LXA}), and $\psi$ and $\bpsi$ transform 
contragrediently:
\bes
\LX \psi &=& -X\psi - D_jX^i\si^j_{ia}T^a\psi, \nle
\LX \bpsi &=& -X\bpsi + D_jX^i\si^j_{ia}\bpsi T^a.
\eens

Clearly, all formulas simplify if we combine spacetime and internal
components into a single $(N+\nm)$-dimensional vector. Let
capital letters from the middle of the alphabeth denote combined
indices, e.g. $I=(\mu,i) = 1, ..., N+n+m$. The Dirac action (\ref{S_D2})
becomes 
\be
S_D &=& \int \dNx\dnmy \bpsi(x,y)\Dslash\psi(x,y), 
\label{S_D3}
\ee
where 
\be
\Dslash \equiv \gm^I(x,y)\DD_I, \qquad
\DD_I = (\dmu,D_i) + A_{aI}(x,y)T^a.
\ee
Given this covariant derivative we can construct the Yang-Mills 
field strength $F_{aIJ}T^a = [\DD_I, \DD_J]$ and the Yang-Mills action
\be
S_{YM} = \int \dNx\dnmy \sqrt{g} g^{IK}g^{JL}F_{aIJ}F^a_{KL}.
\label{S_YM}
\ee
Explicit factors of the metric $g^{IK}$ and the density $\sqrt{g}$ have
been introduced. The metric is defined in terms of the gamma matrices 
$\gm^I(x,y) = (\gm^\mu, \gm^i(x,y))$ as usual:
$\{\gm^I(x,y), \gm^J(x,y)\} = 2g^{IJ}(x,y)$. Note that even if we may 
consider the spacetime components $\gm^\mu$ and $g^{\mu\nu}$ to be
constants (as long as we restrict spacetime diffeomorphisms $\vect(N)$
to its Poincar\'e subalgebra), the internal components $\gm^i(x,y)$
can not be constants because of (\ref{gp2}). We are therefore forced to
consider a curved metric in internal space, even though spacetime
curvature is zero. This issue will not be pursued further here.

A generic conclusion is that {\em we can not consider spacetime to be 
entirely separated from the internal space}; a non-zero value of the 
parameter
$\eps$ in (\ref{gp2}) forces us to combine spacetime and internal indices
into a single entity. However, unlike the situation in presently popular 
theories, {\em spacetime and internal directions are intrinsically 
different from the outset}. 

The action $S = S_D + S_{YM}$ (and perhaps more terms as well) is a 
$\oj$-invariant generalization of the $\oj_0$-invariant standard model
action. However,
if this would be the final goal, passage from $\oj_0$ to the prolong
$\oj = (\oj_-,\oj_0)_*$ would hardly be very interesting, because all
information about $\oj$ tensor modules is already encoded in $\oj_0$.
The genuinely new information in $\oj$ is about form modules, i.e.
tensor modules which are reducible even though the corresponding $\oj_0$
module is irreducible. So we demand that the fermions satisfy not only
the massless Dirac equation 
\be
\Dslash\psi(x,y)=0,
\label{Dir1}
\ee
but also the closedness condition
\be
\nabla\psi(x,y) = 0,
\label{npsi}
\ee
where $\nabla$ is a $\oj$-invariant morphism of the type constructed in
previous sections. It follows from $(\ref{gp2})$ that every such
$\nabla$, which only involves the internal derivative $D_i$, can be 
trivially extended to a $\map(N,\oj)$-invariant morphism.
It is clear that (\ref{npsi}) is compatible with (\ref{Dir1})
and that non-zero solutions exist. Namely, a sufficient condition for
(\ref{npsi}) to hold is $D_i\psi(x,y)=0$, which makes $\psi(x,y)$ 
independent of the internal coordinate $y$. The Dirac equation then
reduces to
\be
\gm^I(x,y)\DD_I\psi(x,y) =
\gm^\mu\DD_\mu\psi(x) = 0,
\label{Dir0}
\ee
which is the usual Dirac equation in flat spacetime. Since any 
solution to (\ref{Dir0}) solves (\ref{Dir1}) and (\ref{npsi}), we 
conclude that non-trivial solutions exist.
					  
However, I have failed to construct an action which produces both
(\ref{Dir1}) and (\ref{npsi}). The closest approximation is obtained
by introducing two Lagrange  multipliers $\la$ and $\chi$ and their
conjugates $\bla$ and $\bchi$.
Assume that there are morphisms $\nabla$ and $\nabla'$:
\be
\ldots \longrightarrow \chi \rarroww{\nabla'} \psi  \rarroww{\nabla} 
\la \longrightarrow +\ldots,
\ee
and duals morphisms, denoted by the same letters:
\be
\ldots \longleftarrow \bchi \larroww{\nabla'} \bpsi  \larroww{\nabla} 
\bla \longleftarrow +\ldots.
\ee
These diagrams indicate how the objects transform; e.g., $\la$ transforms as
$\nabla\psi$, etc.
{F}rom (\ref{npsi}) it follows that $\bpsi$ is also a closed form, i.e.
\be
\nabla'\bpsi = 0.
\ee
The two closedness conditions can now be enforced by the following
Lagrange multiplier action:
\bes
S_{Lag} &=& \int \dNx\dnmy ( \bla\nabla\psi + \nabla'\bpsi\chi)
\nlb{SLag}
&=& - \int \dNx\dnmy ( \nabla\bla\psi + \bpsi\nabla'\chi).
\eens
The Euler-Lagrange equations for the fermions become
\be
\barr{lll}
\Dslash\psi - \nabla'\chi = 0, &\qquad&
\nabla\psi = 0, \\
\DD_I\bpsi\gm^I + \nabla\bla = 0, &\qquad&
\nabla'\bpsi = 0.
\earr
\ee
Since $\psi$ and $\nabla'\chi$ transform in the same way, it is 
tempting to assume that they are proportional. If $m$ is the 
proportionality constant, the equation for $\psi$ becomes
$\Dslash\psi - m\psi = 0$.
Hence the closedness condition (\ref{npsi}) effectively gives rise to
a mass-like term.

It is known that left- and right-handed spinors transform 
differently under internal transformations, so the term
(\ref{SLag}) is in fact not invariant. To remedy this, we need to split 
both $\psi$ and the multipliers into left- and right-handed parts, and
assume that these belong to different form modules. Moreover, we need
spacetime scalars to translate between the two complexes.
These scalars are reminiscent of
Higgs fields, which is natural because they arise for the same reason:
terms like $\bpsi_R\psi_L$ and $\bla_R\nabla\psi_L$ are not invariant.

It is now time to specialize to $\oj_0=\ssg$ and $\oj = \vle(3|6)$
or $\oj = \mb(3|8)$. We first note that the bosons are gauge bosons,
i.e. they are modelled by connections. The $\oj$ connection
$A_{aI}(x,y)T^a$ depends on the internal directions and it has more
components than the $\oj_0$ connection $A_{a\mu}(x)T^a$ has, but it is 
still a $\oj_0$-valued function with a twisted module action;
we still have $a=1,2,...,\dim\oj_0$. Thus, the gauge bosons are 1-1 
with the generators of $\oj_0$. {\em A gauge theory based on $\vle(3|6)$
or $\mb(3|8)$ predicts precisely the 12 gauge bosons of the standard model,
no more and no less:}
\[
\barr{|c|c|c|c|} 
\hline
\hbox{$\oj_0$ weight} & \hbox{Charges} && \hbox{} \\
\hline
(1,1;0;0) & 0 & \hbox{Gluons} & \hbox{$sl(3)$ gauge bosons} \\
(00;2;0) & 1,0,-1 & W^+, W^-, Z & \hbox{$sl(2)$ gauge bosons} \\
(00;0;0) & 0 & \hbox{Photon} & \hbox{$gl(1)$ gauge boson} \\
\hline
\end{array}
\]

We next turn to the assignment of the fundamental fermions, i.e. quarks
and leptons, which we want identify with closed form modules. The 
assignment of $\ssg$ weights to fermions is standard and can be found 
e.g. in \cite{Hua86}, and the corresponding form modules are simply
read off from (\ref{vleforms}) and (\ref{mbforms}).
The electric charge is computed by means of the Gell-Mann-Nishijima 
formula: $Q = I_3+Y/2$, where $I_3$ is the $su(2)$ highest weight and
weak hypercharge $Y = Z/3$ is identified with the
grading operator up to a factor $1/3$.\footnote{My normalization of the
$gl(1)$ eigenvalue differs from Kac and Rudakov \cite{KR00a,KR00b,KR01}
by this factor $1/3$. They want their normalization to agree with 
weak hypercharge in physics, whereas I prefer to use integers throughout.}

\[
\barr{|c|c|ccc|} 
\hline
\hbox{$\oj_0$ weight} & \hbox{Charges} &&&\\
\hline
(0,1;1;1) & {2\/3},-{1\/3} & 
\begin{pmatrix} u_L \\ d_L \end{pmatrix} &
\begin{pmatrix} c_L \\ s_L \end{pmatrix} &
\begin{pmatrix} t_L \\ b_L \end{pmatrix} 
\\
(0,1;0;4) & {2\/3} & u_R & c_R & t_R 
\\
(0,1;0;-2) & -{1\/3} & d_R & s_R & b_R 
\\
\hline
(0,0;1;-3) & 0,-1 & 
\begin{pmatrix} \nu_{eL} \\ e_L \end{pmatrix} &
\begin{pmatrix} \nu_{\mu L} \\ \mu_L \end{pmatrix} &
\begin{pmatrix} \nu_{\tau L} \\ \tau_L \end{pmatrix} 
\\
(0,0;0;-6) & -1 & e_R & \mu_R & \tau_R 
\\
\hline
\hline
(1,0;1;-1) & -{2\/3},{1\/3} & 
\begin{pmatrix} \tilde u_R \\ \tilde d_R \end{pmatrix} &
\begin{pmatrix} \tilde c_R \\ \tilde s_R \end{pmatrix} &
\begin{pmatrix} \tilde t_R \\ \tilde b_R \end{pmatrix} 
\\
(1,0;0;-4) & -{2\/3} & 	\tilde u_L & \tilde c_L & \tilde t_L 
\\
(1,0;0;2) & {1\/3} & \tilde d_L & \tilde s_L & \tilde b_L 
\\
\hline
(0,0;1;3) & 0,1 & 
\begin{pmatrix} \tilde \nu_{eR} \\ \tilde e_R \end{pmatrix} &
\begin{pmatrix} \tilde \nu_{\mu R} \\ \tilde \mu_R \end{pmatrix} &
\begin{pmatrix} \tilde \nu_{\tau R} \\ \tilde \tau_R \end{pmatrix} 
\\
(0,0;0;6) & 1 & \tilde e_L & \tilde \mu_L & \tilde \tau_L 
\\
\hline
\end{array}
\]

\begin{figure}
\begin{center}
\includegraphics{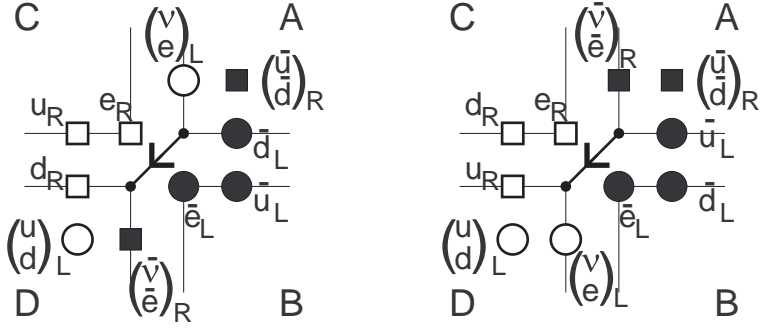}
  \caption{Assignment of first-generation fermions for $\vle(3|6)$ (left)
  and $\mb(3|8)$ (right). White = particles and black = anti-particles.
  Circles = left-handed and squares = right-handed.
  The arrows in the middle indicate the overall direction of the
  morphisms, and the sectors are labelled as in Figure \ref{fig:vle}.
}
    \label{fig:vpart}
  \end{center}
\end{figure}

The best assignment possible is described by the following table:
\[
\barr{|c|c|c|c|c|c|c|} 
\hline
\hbox{$\oj_0$ weight} & &
z & \hbox{$\vle$ form}& z(\vle)&\hbox{$\mb$ form}& z(\mb) \\
\hline
(0,1;1;1) & \begin{pmatrix} u_L \\ d_L \end{pmatrix} &
1 & \Om_D(1,1) & 1 & \Om_D(1,1) & 1
\\
(0,1;0;4) &  u_R & 
4 & \Om_C(1,0) & -8 & \Om_D(1,0) & 4
\\
(0,1;0;-2) & d_R & 
-2 & \Om_D(1,0) & -2 & \Om_C(1,0) & 10
\\
\hline
(0,0;1;-3) & \begin{pmatrix} \nu_{eL} \\ e_L \end{pmatrix} &
-3 & \Om_A(0,1) & -3 & \Om_D(0,1) & -3
\\
(0,0;0;-6) & e_R &
-6 & \Om_C(0,0) & -6 & \Om_C(0,0) & 6
\\
\hline
\hline
(1,0;1;-1) & \begin{pmatrix} \tilde u_R \\ \tilde d_R \end{pmatrix} &
-1 & \Om_A(1,1) & -1 & \Om_A(1,1) & -7
\\
(1,0;0;-4) & \tilde u_L &
-4 & \Om_B(1,0) & 8 & \Om_A(1,0) & -10
\\
(1,0;0;2) & \tilde d_L &
2 & \Om_A(1,0) & 2 & \Om_B(1,0) & -16
\\
\hline
(0,0;1;3) & \begin{pmatrix} \tilde \nu_{eR} \\ \tilde e_R \end{pmatrix} &
3 & \Om_D(0,1) & 3 & \Om_A(0,1) & -3
\\
(0,0;0;6)  & \tilde e_L & 
6 & \Om_B(0,0) & 6 & \Om_B(0,0) & -12
\\
\hline
\end{array}
\]
The assignment of fermions is illustrated in Figure \ref{fig:vpart}.

It is quite remarkable that
it is possible to fit the $sl(3)\ooplus sl(2)$ weights of all 
fundamental fermions so snugly in the list of form modules. It is e.g.
not at all possible to identify the $sl(3)$ gauge bosons with forms,
due to the $(1,1)$ $sl(3)$ weight. On other hand, we know that the
gauge bosons must be connections, so identifying them with form modules
is out of the question anyway. 
However, not all $gl(1)$ weights come out correctly. $\vle(3|6)$ predicts
that the right-handed u quark $u_R$ should have $z=-8$, which disagrees
with the correct value $z=4$ $(Y=4/3)$ by two units of hypercharge. 
Similarly, the left-handed u anti-quark $\tilde u_L$ has $z=8$ versus 
the correct $z=-4$. 
For $\mb(3|8)$ the discrepancy is even bigger, but here the error in
$z$ is constant in each of the four sectors:
\[
\barr{c|cccc}
& A & B & C & D \\
\hline
\Delta z & -6 & -18 & +12 & 0\\
\earr
\]
Upon second quantization, the $\oj_0$-invariant fermions in the standard 
model become field operators which satisfy the Clifford algebra
\be
\{\bpsi(x),\psi(x')\} = \dlt^N(x-x').
\label{Cl0}
\ee
Similarly, we assume that the $\oj$-invariant fermions satisfy the 
Clifford algebra
\be
\{\bpsi(x,y),\psi(x',y')\} = \dlt^N(x-x')\dlt^\nm(y-y').
\label{Cl1}
\ee
Elimination of the second delta function requires precisely one 
integration over internal space. The simplest way to relate
(\ref{Cl0}) and (\ref{Cl1}) is to put
\be
\psi(x) = \psi(x,y_0), \qquad \bpsi(x) = \int \dnmy \bpsi(x,y),
\label{projg}
\ee
where $y_0$ is a constant, because $\int\dnmy\dlt^\nm(y-y_0) = 1$.
For $\mb(3|8)$, this integral adds $+6$ to the weight for the 
anti-fermions, which leaves us with the following errors:
\be
\barr{c|cccc}
& A & B & C & D \\
\hline
\Delta z & 0 & -12 & +12 & 0\\
\earr
\label{mberr}
\ee
For $\vle(3|6)$ the volume form has weight zero ($\vle(3|6)$ vector 
fields are divergence free), so integration does not change hypercharge.

It is more difficult to reconcile the values of $z$ in sector C,
but this problem could be related to the ``wedge product'', i.e. the
mapping of a pair of  form modules into a new one. In the D sector,
we can define a map in the natural way:
\bes
\wedge: &&\Om_D(q,r)\otimes\Om_D(q',r') 
\longrightarrow \Om_D(q+q',r+r') \nl
&&f(y)\bgm_{i_1a_1}...\bgm_{i_{q}a_{q}}\bbt_{a_{q+1}}...\bbt_{a_r} \otimes
g(y)\bgm_{j_1b_1}...\bgm_{j_{q'}b_{q'}}\bbt_{b_{q'+1}}...\bbt_{b_{r'}}
\mapsto \nl 
&&\mapsto f(y)g(y)
\bgm_{i_1a_1}...\bgm_{i_{q}a_{q}}\bgm_{j_1b_1}...\bgm_{j_{q'}b_{q'}}
\bbt_{a_{q+1}}...\bbt_{a_r}\bbt_{b_{q'+1}}...\bbt_{b_{r'}}.
\eens
In terms of $\oj_0$ weights,
\bes
\wedge:&&(0,q;r;4q-3r) \otimes (0,q';r';4q'-3r') \to\nle
&&\to (0,q+q';r+r';4(q+q')-3(r'+r')).
\eens
However, the wedge product in the C sector must be defined by
\bes
\wedge:&&(0,q;r;4q+3r+6) \otimes (0,q';r';4q'+3r'+6) \to\nle
&&\to (0,q+q';r+r';4(q+q')+3(r'+r')+6).
\eens
The total value of $z$ is $(4q+3r+6) + (4q'+3r'+6)$ to the left but
only $4(q+q')+3(r'+r')+6$ to the right; six units of $z$ (i.e.
two units of hypercharge) have been ``eaten'' by the wedge product.
A reasonable conclusion is that the interpretation of the value of $z$ 
in the C sector is unclear. It might be possible to add
an overall multiple of $6$, but this multiple should be the same for all
modules in the sector. After addition of $\Delta z = -12$ to the C sector
(and $\Delta z = +12$ to the B sector), the hypercharge assignments 
agree with the standard model.

Suitable wedge products can be defined in other sectors similarly; the 
wedge product between forms from different sectors involves contraction of
dual indices. The wedge product appears in Figure \ref{fig:vle} simply
as vector addition; associativity and graded commutativity follows 
immediately because vector addition has these properties.
There is some ambiguity when the results ends on some of the coordinate 
axes, since then there are two different form modules with different value
of $z$. It is remarkable that {\em the fundamental fermions 
in the first generation are exactly right to make it possible to build 
\underline{all} form modules from wedge products}. It might appear from 
Figure \ref{fig:vpart} that there are too many fermions, but we must
keep in mind that the fermions are spinors as well, so there is an extra 
quantum number due to spin.

A $\oj_0$ gauge theory is symmetric under interchange of particles and
anti-particles, but not the $\oj$ theory. In Figure \ref{fig:vpart}, CP
amounts to a reflection in the origin, but the direction of the morphisms
(indicated by the diagonal arrow) is unchanged. To obtain complete
reflection, we must also reverse the direction of all morphisms, which
can be interpreted as time reversal T. Thus CPT is conserved, at least
in this figure\footnote{Kac and Rudakov \cite{KR01} claim that CPT is 
violated for $\mb(3|8)$, but the relevance to the present 
model is unclear to me. Of course, $\psi$ and $\bpsi$ are treated 
completely  differently in (\ref{projg}), but this equation is not 
intrinsic; its purpose is to establish a link to the usual standard 
model.}, but CP and T are manifestly broken. Note the difference
to the standard model, where breaking of CP is dynamical but not
manifest. The inequivalence of dual modules is completely analogous to
the difference between contravariant and covariant tensors: conventional
in $gl(N)$, but substantial in $\vect(N)$. Only covariant skew tensor
fields admit a morphism, the exterior derivative.

Despite the fact that $\vle(3|6)$ and $\mb(3|8)$ are superalgebras, it 
seems that {\em no supersymmetric partners arise in the second-gauged 
theories}. The multiplets have both bosonic and fermionic components, but
the weights are the same as for $\ssg$ and only the lowest-weight
states are identified with particles. However, this point requires 
further study.

We have identified fermions in the first generation with form modules.
What about the second and third generations? Since later generations
have the same quantum numbers as the first, the natural thing is to
combine several generations, i.e. $\oj_0$ modules, into a single $\oj$
module. However, the decomposition of the $\oj$ form modules into 
irreducible $\oj_0$ modules has not yet been studied.

All considerations in this paper are classical. Quantization of
a second-gauged theory will presumably encounter problems similar to those
in quantum gravity. On the mathematical side, the construction of Fock
modules with the normal ordering prescription gives rise to a
higher-dimensional generalization of the Virasoro algebra; such modules 
were first constructed for $\vect(N)$ by Rao and Moody \cite{RM94}
and in the super case by myself \cite{Lar97}. It is clear that every
vectorial Lie superalgebra $\oj \subset \vect(n|m)$ have similar Fock
representations; the geometrical construction in \cite{Lar98} goes 
through also for superalgebras. However, ordinary quantization methods
only work when the symmetry algebras have no or possibly central 
extensions, and the Virasoro-like cocycle is non-central except for
algebras of linear growth.

We conclude with a summary of the main features and predictions of the 
second-gauged standard model based on $\mb(3|8)$:
\begin{enumerate}
\item
All fermions in the first generation can be identified with closed form 
modules, but the assignment of hypercharge is unclear, maybe wrong.
\item
All closed form modules can be built by taking suitably defined wedge 
products of the fundamental fermions.
\item
The gauge bosons acquire more components, but there are still only
$\dim\oj_0 = 12$ different types.
\item
There are presumably no supersymmetric partners, since irreducible 
$\oj$ modules, and hence fundamental particles, are labelled by $\oj_0$ 
weights.
\item
The symmetry between particles and anti-particles is manifestly broken.
\item
Several generations might arise by decomposing $\oj$ irreps into $\oj_0$
irreps.
\item
The internal space must be combined with spacetime, but internal 
directions are nevertheless fundamentally different from the spacetime 
directions.
\end{enumerate}
Although there are shortcomings, in particular concerning the unclear 
assignment of hypercharge, I believe that this list is sufficiently close
to experimental data to merit further study of the 
second-gauged standard model.

\end{document}